\documentclass[aip,preprint,letterpaper,groupedaddress,longbibliography]{revtex4-2}
\usepackage{amsmath,amsthm,amsfonts}
\theoremstyle{definition}

\usepackage{multirow}
\usepackage{array}
\usepackage{soul}
\usepackage{float}
\usepackage{graphicx}
\usepackage{afterpage}
\usepackage{tabularx}

\usepackage[font=small,labelsep=quad,format=plain]{caption}
\usepackage{subcaption}
\usepackage[colorlinks=true,citecolor=blue,linkcolor=blue,urlcolor=blue]{hyperref}
\usepackage{cleveref}
\crefname{equation}{Eq.}{Eqs.}
\crefname{section}{Sec.}{Secs.}
\crefname{subsection}{Sec.}{Secs.}
\crefname{appendix}{Appendix}{Appendices}
\crefname{figure}{Fig.}{Figs.}
\crefname{table}{Table}{Tables}
\crefname{proposition}{}{}
\crefname{corollary}{}{}

\usepackage{bm}
\usepackage[dvipsnames]{xcolor} 

\usepackage{siunitx}

\begin{document}

\newcommand{\sio}[1]{SiO\textsubscript{2}}
\newcommand{\h}[1]{H\textsubscript{2}}
\renewcommand{\sf}[1]{SF\textsubscript{6}}
\renewcommand{\o}[1]{O\textsubscript{2}}

\title{Atomic layer etching of \texorpdfstring{SiO\textsubscript{2} using sequential exposures of Al(CH\textsubscript{3})\textsubscript{3} and H\textsubscript{2}/SF\textsubscript{6} plasma}{SiO2 using sequential exposures of Al(CH3)3 and H2/SF6 plasma}}

\author{David S. Catherall}
\affiliation{Division of Engineering and Applied Science, California Institute of Technology, Pasadena, CA 91125, USA}
\author{Azmain A. Hossain}
\affiliation{Division of Engineering and Applied Science, California Institute of Technology, Pasadena, CA 91125, USA}
\author{Austin J. Minnich}
\email{aminnich@caltech.edu}
\affiliation{Division of Engineering and Applied Science, California Institute of Technology, Pasadena, CA 91125, USA}
\date{\today}

\begin{abstract}
On-chip photonic devices based on \sio2 are of interest for applications such as microresonator gyroscopes and microwave sources. Although \sio2 microdisk resonators have achieved quality factors exceeding one billion, this value remains an order of magnitude less than the intrinsic limit due to surface roughness scattering. Atomic layer etching (ALE) has potential to mitigate this scattering because of its ability to smooth surfaces to sub-nanometer length scales. While isotropic ALE processes for \sio2 have been reported, they are not generally compatible with commercial reactors, and the effect on surface roughness has not been studied. Here, we report an ALE process for \sio2 using sequential exposures of Al(CH\textsubscript{3})\textsubscript{3} (trimethylaluminum, TMA) and Ar/\h2/\sf6 plasma. We find that each process step is self-limiting, and that the overall process exhibits a synergy of 100\%.  We observe etch rates up to 0.58 \AA~per cycle for thermally-grown \sio2 and higher rates for ALD, PECVD, and sputtered \sio2 up to 2.38 \AA~per cycle. Furthermore, we observe a decrease in surface roughness by 62\% on a roughened film. The residual concentration of Al and F is around 1-2\%, which can be further decreased by \o2 plasma treatment. This process could find applications in smoothing of \sio2 optical devices and thereby enabling device quality factors to approach limits set by intrinsic dissipation.

\end{abstract}

\maketitle
\newpage

\section{Introduction}

Atomic layer etching (ALE) is a subtractive nanofabrication process of increasing interest due its atomic-scale etch depth control and selectivity \cite{Oehrlein:2015,Fang:2018,George:2020,Lill:2021,Fischer:2021,Fischer:2023}, as well as its ability to smooth surfaces down to the sub-nanometer scale \cite{Kanarik:2018,Gerritsen:2022,Chen:2022}. In contrast to continuous etching processes such as reactive ion etching (RIE), ALE consists of two or more self-limiting surface treatments \cite{Oehrlein:2015,Fang:2018,George:2020}. In the first step, the surface of a material is chemically modified by exposure to a substance such as a gas, plasma, or organometallic vapor. This modification is self-limiting and affects only the surface region. A second step volatilizes the modified material by chemical reaction, thermal cycling, or ion impingement. These steps can then be repeated as many times as desired. The number of known ALE processes has grown substantially over the last decade and now includes metals (including Co \cite{Pacco:2019}, Cu \cite{Gong:2018,Mohimi:2018,Sheil:2021}, and W \cite{Johnson:2017,Xie:2018}), semiconductors (including Si \cite{Athavale:1996,Park:2005}, InP \cite{Park:2006,Ko:1993}, and GaAs \cite{Aoyagi:1992,Meguro:1993,Ko:1993}), dielectrics (including Al\textsubscript{2}O\textsubscript{3} \cite{Min:2013,Lee:2016,Hennessy:2017,Chittock:2020}, \sio2 \cite{Lee:2016,Metzler:2013,Kaler:2017,Koh:2017}, and HfO\textsubscript{2} \cite{Lee:2016,Lin:2020}), and a superconductor (TiN \cite{Azmain:2023}).

Etching methods for SiO\textsubscript{2} are of particular interest in nanofabrication owing to the presence of \sio2 in semiconductor, superconductor, and optical devices. Processing of \sio2 with low damage and surface roughness is especially critical when it is the active material in a device, as occurs in on-chip photonic devices such as microcomb systems, microresonator gyroscopes, and microwave sources \cite{Yi:2015,Wu:2020}. These devices all rely on resonators such as silica microresonators which now exhibit quality factors ($Q$) as high as $10^9$ (Refs. \onlinecite{Wu:2020,Wu:2023}), where $Q$ is defined as the ratio of center resonance frequency to the resonance bandwidth. The maximum theoretical $Q$ for these microresonators is $\sim 8\times 10^9$, however, and the difference between this maximum value and the lesser value achieved in practice has been attributed primarily to surface roughness created by fabrication processes \cite{Lee:2012,Wu:2023}. Therefore, to increase the $Q$, an etch process capable of smoothing to sub-nanometer length scales is needed. ALE has potential to fulfill this need; although ALE etches too slowly to be used as the primary processing etch, it could be used to increase the $Q$ of these devices as a post-processing smoothing step.


There are two currently known classes of isotropic ALE processes for \sio2. In the first process class, the \sio2 surface is modified by conversion to ammonium fluorosilicate ((NH\textsubscript{4})\textsubscript{2}SiF\textsubscript{6}) using a combination of NH\textsubscript{3} and fluorination agent, such as CF\textsubscript{4} plasma \cite{Cho:2020}, NF\textsubscript{3} plasma \cite{Cho:2020}, HF vapor \cite{Miyoshi:2021,Ohtake:2023}, \h2/NF\textsubscript{3} plasma \cite{Gill:2021,Gil:2023}, or \h2/\sf6 plasma \cite{Miyoshi:2021}. This step is followed by desorption via heating, either using halogen lamps \cite{Cho:2020}, infrared lamps \cite{Miyoshi:2021,Ohtake:2023}, or a vacuum break and a hot plate \cite{Gill:2021,Gil:2023}. This approach yields etch rates in the range of $10-100$ \AA/cycle. The second class of processes uses a ``conversion etch'' whereby \sio2 is modified by conversion to alumina and aluminosilicates using trimethylaluminum (TMA) exposures, followed by fluorination induced by HF vapor \cite{DuMont:2017,Rahman:2018} or CHF\textsubscript{3} \cite{Rahman:2018}, which forms fluorinated alumina and aluminosilicate compounds. Another TMA exposure causes transmetalation and volatilization of the modified surface, while simultaneously forming more alumina and aluminosilicates. This approach yields etch rates of $\sim 0.31-0.35$ \AA/cycle.

Each method has practical complications when applied to device fabrication. These issues include the lack of atomic scale control (etch rates are $> 5$ \AA/cycle) \cite{Cho:2020,Miyoshi:2021,Gill:2021,Gil:2023,Ohtake:2023}; long cycle times (from 155 s to over 10 min); inconsistent etch rates \cite{Rahman:2018}; the use of HF vapor \cite{Miyoshi:2021,Ohtake:2023,DuMont:2017,Rahman:2018}, which cannot be used in some commercial ALD systems; high pressures ($> 250$ mTorr) \cite{DuMont:2017,Rahman:2018}, which is outside the process range for some commercial ALD systems; the use of lamp heating \cite{Cho:2020,Miyoshi:2021,Ohtake:2023}, which requires special system designs and increases cycle time; or breaking vacuum \cite{Gill:2021,Gil:2023}. Additionally, the only measurements of surface roughness show either no effect \cite{Gil:2023} or an increase after ALE \cite{Gill:2021}. Therefore, a new process that is compatible with commercial tools is needed to facilitate the application of ALE for smoothing of \sio2 optical microdevices.

Here, we report an ALE process for \sio2 using sequential exposures of TMA and an Ar/\h2/\sf6 plasma. We demonstrate that a sufficient hydrogen content in an Ar/\sf6 plasma reduces the spontaneous etch of \sio2 to negligible levels while sufficiently modifying the surface for ALE. We observe an etch-per-cycle (EPC) at \qty{400}{\degreeCelsius} of up to 0.58 \AA~on thermally oxidized \sio2, and up to 2.38 \AA~on sputtered \sio2. This process exhibits 100\% synergy and self-limiting characteristics. We performed surface roughness measurements after ALE and observed a 62\% decrease in the roughness of an RIE-roughened surface. Furthermore, XPS reveals less than 2\% combined F and Al contamination post-ALE, nearly an order of magnitude less than other reports. Our work paves the way for surface smoothing of \sio2 microdevices using widely-available processing tools, potentially leading to on-chip photonic devices limited only by intrinsic dissipation.

\section{Methods} \label{sec:methods}

All processes were conducted in an Oxford FlexAL II system unless otherwise stated. The process chamber was preconditioned by an Ar/\o2/\sf6 plasma clean (40 sccm \sf6, 40 sccm \o2, 120 sccm Ar) for 6 minutes to remove contaminants with volatile fluorides. This cleaning was followed by 300 cycles of Al\textsubscript{2}O\textsubscript{3} ALD per four hours of ALE to coat the chamber in a material with a nonvolatile fluoride. Next, the ICP and chamber were conditioned to the ALE plasma by running a 2 minute Ar/\h2/\sf6 plasma (42 sccm \h2, 8 sccm \sf6, 150 sccm Ar). Lastly, 10 cycles of ALE were performed with the carrier wafer in the chamber. Samples were then inserted into the reactor for processing. Except where otherwise specified, the table temperature was \qty{400}{\degreeCelsius}, and plasma steps were operated with an ICP power of 100 W at a pressure of 100 mTorr.

Our ALE process is motivated by ALE processes for \sio2 based on HF and TMA exposures. However, HF vapor is incompatible with the Oxford FlexAL tool. Therefore, we employed an Ar/\h2/\sf6 plasma instead following recent works \cite{Chittock:2020,Azmain:2023,Miyoshi:2021,Gill:2021,Gil:2023}. The inclusion of \h2 has been found to produce HF \textit{in-situ} \cite{Volynets:2020,Pankratiev:2020} which functions in the etching process similarly to HF delivered from a vapor source.

The ALE process was performed using the following sequence of steps. First, all valves were closed except the automatic pressure control valve (APC), the chamber was pumped down for 5 s, the APC valve was closed, and an initial TMA pulse of 610 ms was used to bring the chamber to $\sim 200 \pm10$ mTorr. After 390 ms, another 10 ms TMA pulse was admitted, and this boosting pulse was repeated every 2 seconds thereafter until the total TMA dose time was 51 seconds. The TMA dose was followed by a 2 s purge step with Ar. The TMA purge was followed by a plasma exposure of 42 sccm \h2, 8 sccm \sf6, and 150 sccm Ar (gases stabilized for 3 seconds before striking). The plasma dose time was 15 s. The Ar/\h2/\sf6 plasma mixture is hereafter referred to as ``\textit{in-situ} HF". The plasma exposure was followed by another 2 s purge to end the cycle. No change in etch rates was observed for purge times ranging from 2-10 s, and 2 s was chosen to minimize the cycle time. The overall process was terminated with a final TMA step.

Except where specified, \sio2 thin film samples used for ALE experiments were prepared by dry-oxidizing high-resistivity ($> 20$ k$\Omega$cm) prime-grade (100) silicon wafer (University Wafer) to a measured oxide thickness of $70-80$ nm. Film thicknesses were measured via \textit{ex-situ} ellipsometery using a J.A. Woolam M-2000 spectroscopic ellipsometer. Measurements were taken in a $3\times3$ grid across the center $10\times10$ mm of a $13\times13$ mm sample. Each measurement was taken from $370-1000$ nm, at 60\textdegree, 65\textdegree, and 70\textdegree, and using a dwell time of 2 s. No focusing probes were used, resulting in a beam diameter of $\approx 3$ mm. The thickness of \sio2 was determined using a fit to a film stack consisting of Si wafer, Si/\sio2 interface, and thermal \sio2 film, with all model data from Ref.~\onlinecite{Herzinger:1998}. Samples were oriented in the same position for each scan, and the uncertainty is taken as the sum of the fit error as reported by the ellipsometer and two standard deviations of the distribution of etch depths measured across the nine points.

Surface mapping was conducted using a Bruker Dimension Icon atomic force microscope (AFM) in PeakForce Tapping mode using a ScanAsyst-Air AFM probe. Two types of scans were taken: $500 \times 500$ nm with a scan rate of 0.5 Hz and 256 samples per line, and $1000 \times 20$ nm with a scan rate of 0.1 Hz and 1024 samples per line. Post-processing was performed on the $500 \times 500$ nm scans via the Bruker Analysis flatten operation at first order, and the roughness parameters $R_q$ and $R_a$ (defined as the root mean square and average deviations, respectively) were then taken from the resulting surface map. The other scans were used for a surface noise analysis described in \cref{sec:roughness}.

XPS was performed using a Kratos Axis Ultra X-ray photoelectron spectrometer and monochromatic Al K$\alpha$ source. XPS data was analyzed using CASA-XPS (Casa Software Ltd.) with a U 2 Tougaard background. RSF values were taken from the internal CASA library for Kratos tools.


\section{Results}

\subsection{\texorpdfstring{SF\textsubscript{6} to \h2 flow rate ratio and \sio2 spontaneous etching}{SF6 to H2 flow rate ratio and SiO2 spontaneous etching}}

\begin{figure}
    \centering{
        \includegraphics[width=3.4in, height=3.0in]{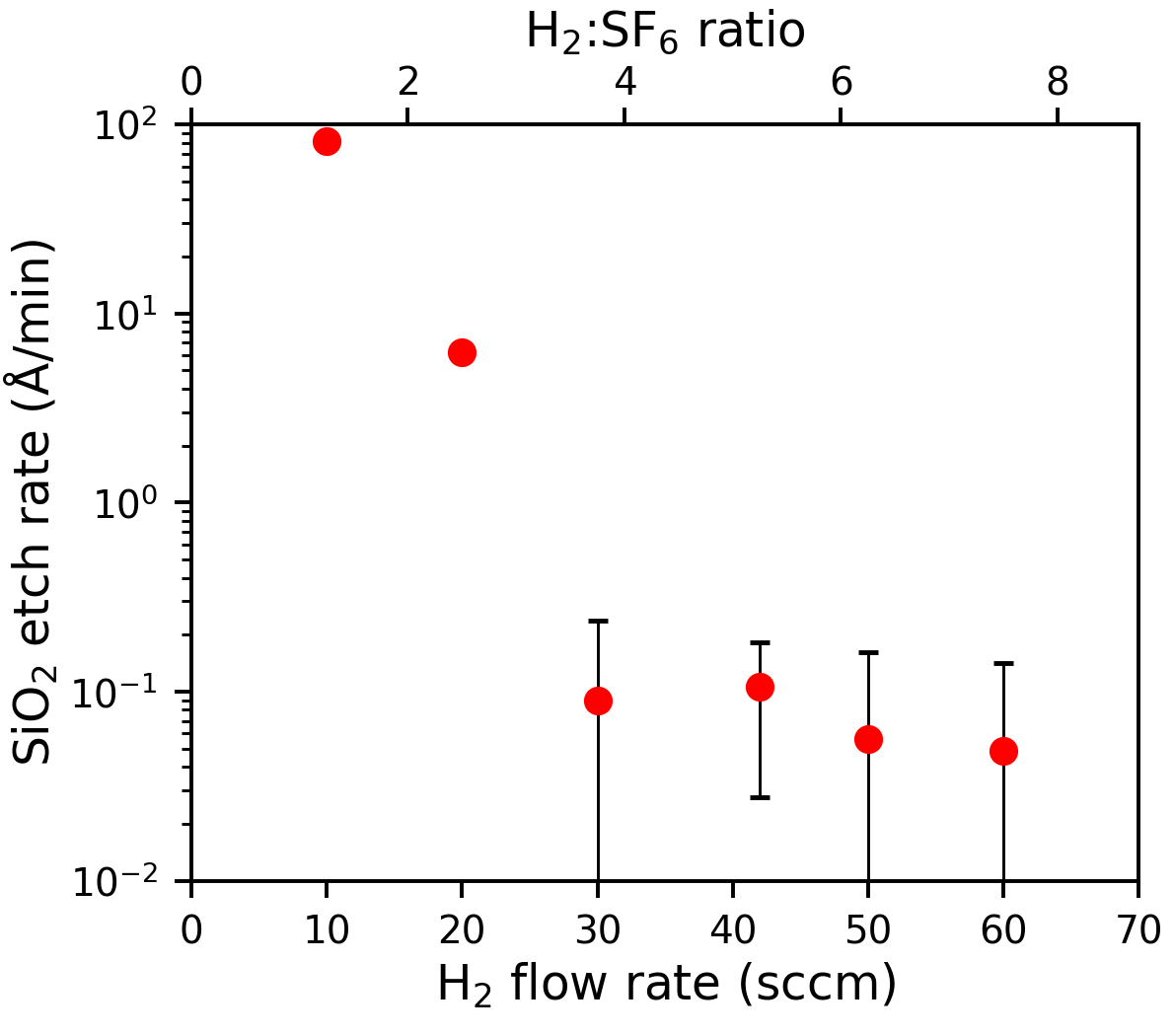}}
    \caption{Spontaneous etch rate of \sio2 when exposed to an Ar/\h2/\sf6 plasma versus \h2 flow rate. Ar and \sf6 flow rates were held constant at 150 and 8 sccm, respectively. The point at 42 sccm \h2 is referred to throughout this work as ``\textit{in-situ} HF''.}
    \label{fig:spont_etch}
\end{figure}

We begin by characterizing the spontaneous etch rate of \sio2 versus the hydrogen flow rate into an Ar/\sf6 plasma, to determine the \h2 flow rate which eliminates spontaneous etching and thereby facilitates ALE. An Ar/\h2/\sf6 plasma has been used for ALE of \sio2 previously, but the effect of the plasma exposure half-cycle was not reported \cite{Miyoshi:2021}. The spontaneous etch rate of \sio2 in Ar/\sf6 was measured by exposing \sio2 samples to plasmas with varying \h2 flow rate. The Ar and \sf6 flow rates were 150 and 8 sccm, respectively. For these experiments, samples were pre-treated by a 2 min 150 sccm Ar, 35 sccm \h2, 15 sccm \sf6 plasma, followed by a 1 min 150 sccm Ar, 42 sccm \h2, 8 sccm \sf6 plasma to begin etching into the bulk and clean the surface.

The measured etch rate versus \h2 flow rate is shown in \cref{fig:spont_etch}. The data indicate that for \h2 flow rates at or exceeding 30 sccm, corresponding to an \h2:\sf6 flow rate ratio $\geq4$, the etch rate is negligible. This result indicates that it is possible to quench a sufficient quantity of the fluorine radicals in an Ar/\sf6 plasma with hydrogen such that the primary fluorination agent becomes HF \cite{Pankratiev:2020,Volynets:2020}, rendering spontaneous etching of \sio2 negligible. The HF formed \textit{in-situ} can then be used for ALE.

\subsection{Synergistic etching and saturation curves}

\begin{figure}
    \centering{
        \includegraphics[width=3.4in, height=3.0in]{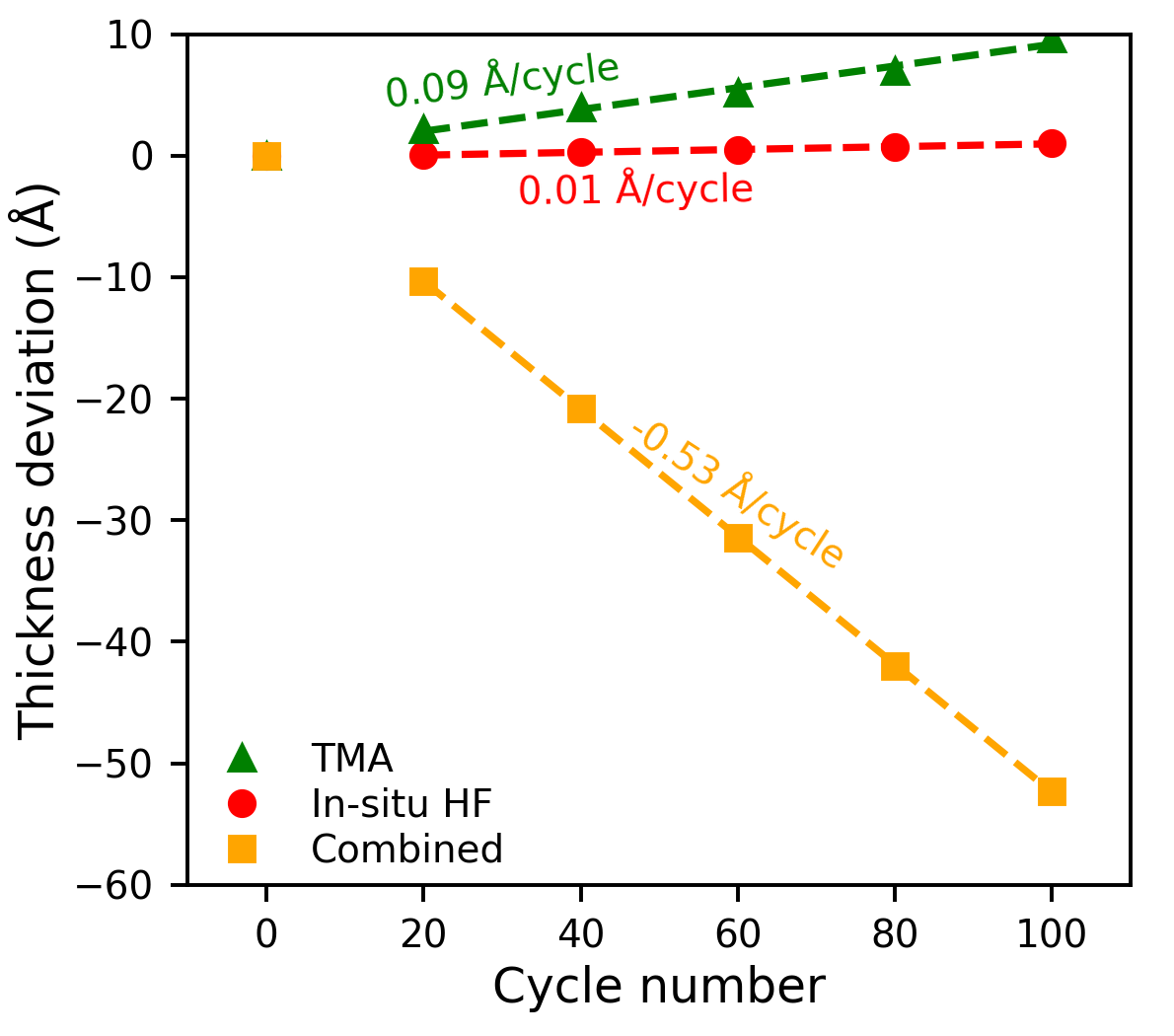}}
    \caption{Thickness change versus cycle number using only \textit{in-situ} HF (red circles), only TMA (green triangles), and both half-cycles together (gold squares) measured using ex-situ ellipsometery. Dashed lines of the same color are linear fits to the points from $20-100$ cycles. The per-cycle thickness changes for TMA only, \textit{in-situ} HF only, and both half-cycles are 0.09, 0.01, and -0.53 \AA~/cycle, respectively.}
    \label{fig:half_cycles}
\end{figure}

We next show the thickness change of \sio2 after repetitions of each half cycle (TMA and \textit{in-situ} HF) as well as the full ALE process in \cref{fig:half_cycles}. Because samples were not preconditioned, the first 20 cycles are omitted from the analysis to account for any non-steady-state etching. First, with only TMA half cycle exposures, we observe an increase in film thickness of $\approx 0.09$ \AA~/cycle. This increase could be attributed to the formation of alumina and aluminosilicates due to TMA reacting with \sio2, and adsorbed aluminum species oxidizing during ex-situ ellipsometery. With only \textit{in-situ} HF exposures, we observe a negligible change in the film thickness. With both half cycles, we observe a linear ($R^2 = 0.99997$) decrease in the film thickness with an etch rate of 0.53 \AA~/cycle. Because neither half cycle results in etching, the recipe exhibits a synergy of 100\%.

\begin{figure}
    \centering{
        \phantomsubcaption\label{subfig:sat_a}
        \phantomsubcaption\label{subfig:sat_b}
        \includegraphics[width=3.4in, height=5in]{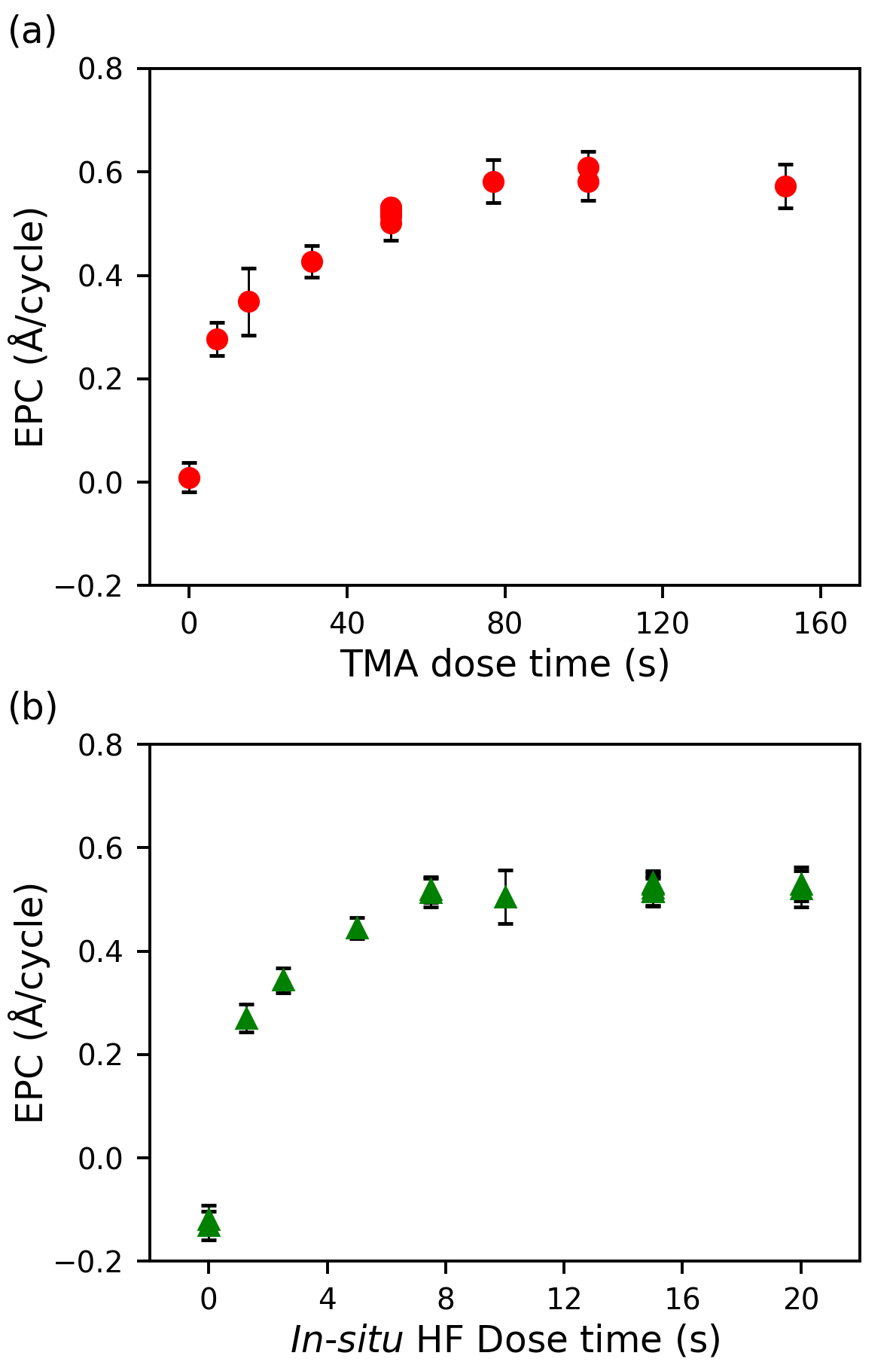}}
    \caption{(a) Etch per cycle versus TMA dose time with \textit{in-situ} HF exposure time fixed at 15 s. (b) Etch per cycle versus \textit{in-situ} HF dose time when the TMA dose was fixed at 51 s.}
    \label{fig:saturation}
\end{figure}

To ensure that each ALE half cycle is self-limiting, we measured the EPC versus dose time and show the data in \cref{fig:saturation}. When the TMA exposure time is varied independent of the \textit{in-situ} HF time (constant at 15 s), saturation occurs after $\sim80$ s as shown in \cref{subfig:sat_a}. When the \textit{in-situ} HF exposure time is varied while the TMA time is kept constant (51 s), saturation is observed after $\sim 8$ s. In this latter case, we also find deposition occurs when the dose time is zero, possibly due to the formation of alumina and aluminosilicates. The fully saturated etch rate was found to be $\sim 0.58$ \AA~per cycle.

\subsection{EPC variation with temperature and pressure}

\begin{figure}
    \centering{
        \phantomsubcaption\label{subfig:temp_press_a}
        \phantomsubcaption\label{subfig:temp_press_b}}
        \includegraphics[width=3.4in, height=5.2in]{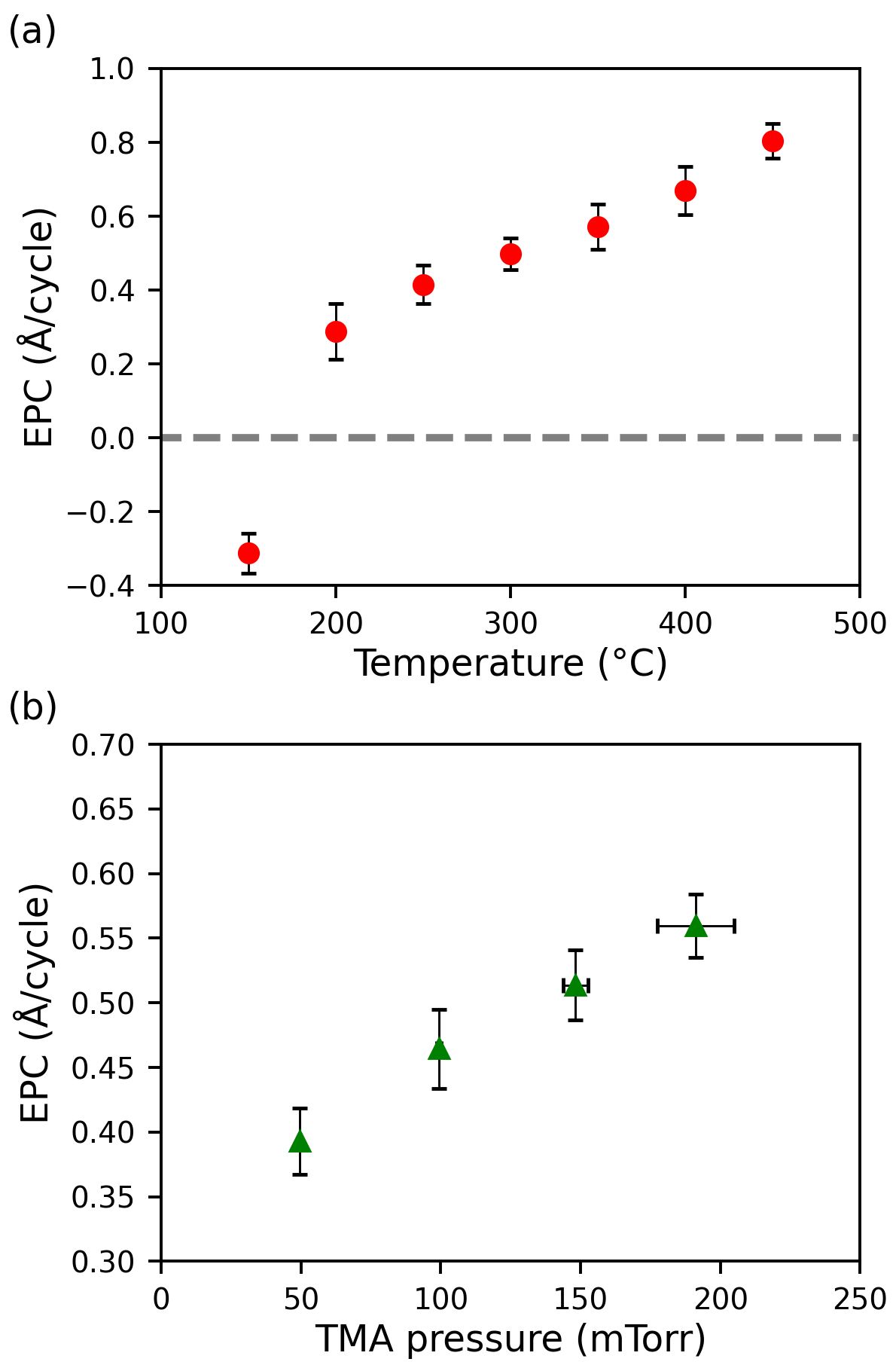}
    \caption{(a) Etch per cycle versus table temperature. At \qty{150}{\degreeCelsius} deposition was observed, while at \qty{200}{\degreeCelsius} and higher etching was observed. Etch rates are larger than reported in other figures due to samples exposure to $\sim 24$ hours of atmosphere between preconditioning and the experiment. (b) Etch per cycle versus TMA pressure. Horizontal error represents the two-sigma deviation from the mean pressure during the experiment.}
    \label{fig:temp_press}
\end{figure}

Next, we present the dependence of EPC on table temperature in \cref{subfig:temp_press_a}. Temperature was measured by a thermocouple integrated into the sample stage. Samples were preconditioned with 10 cycles of ALE $\approx 24$ hours prior to usage. We observe that at \qty{150}{\degreeCelsius} deposition occurs, possibly due to the formation of AlF\textsubscript{3} as was reported in a previous ALE study of alumina \cite{Chittock:2020}. At \qty{\geq200}{\degreeCelsius} etching occurs, and the EPC increases approximately linearly with increasing temperature.


We additionally show the dependence of EPC on TMA pressure in \cref{subfig:temp_press_b}. In these experiments the TMA dose time was fixed at 51 s. We observe a linear trend in the range measured, although it is expected that the etch rate decreases to zero at zero pressure to be consistent with the findings in \cref{fig:half_cycles,fig:saturation}. The EPC does not plateau for pressures up to 250 mTorr, which suggests that reactors capable of higher pressures could achieve either faster EPC saturation or a larger EPC above 250 mTorr. These findings differ from prior HF/TMA \sio2 ALE works as we observe significant etching ($> 0.2$ \AA/cycle) at pressures below 800 mTorr \cite{DuMont:2017,Rahman:2018}.

\subsection{Post-ALE surface composition}

\begin{figure}
    \centering{
        \includegraphics[width=7in, height=3.5in]{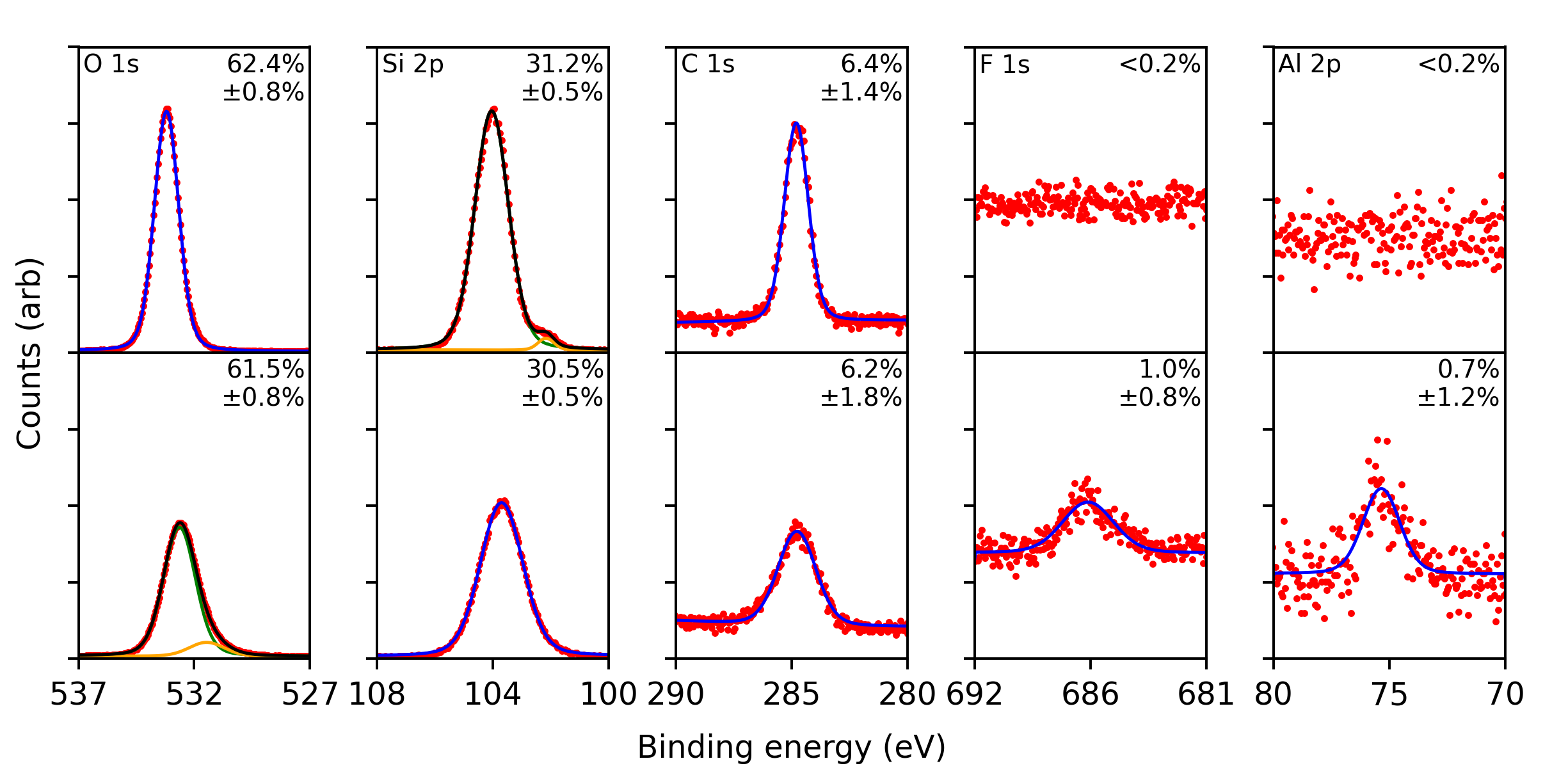}}
    \caption{Surface XPS data for an untreated (upper plots) and ALE treated (lower plots) \sio2 film. From left to right,  peaks correspond to O 1s, Si 2p, C 1s, F 1s, and Al 2p. Sulfur not shown as it was not detected. Experimental data shown as red points, and blue lines represent the peak fit for single peaks. Where two sub-peaks are present they are shown as green and yellow lines, with a black envelope.}
    \label{fig:xps}
\end{figure}

We used surface XPS to characterize the composition of three different samples: untreated \sio2, ALE treated \sio2 (300 cycles of ``rapid'' ALE described in \cref{sec:roughness}), and ALE treated \sio2 with a post-process O\textsubscript{2} plasma (5 min, 60 sccm O\textsubscript{2}, \qty{400}{\degreeCelsius}, 15 mTorr, 400 W ICP power). The post-processing step on the last sample was done in an effort to replace surface fluorine species with oxides, decreasing surface contamination. The XPS spectra are shown in \cref{fig:xps} and film compositions are tabulated in \cref{tab:XPS_data}. Spectra for the oxygen plasma-treated sample are qualitatively similar to those of the ALE-treated sample (not shown). For the untreated film, oxygen and silicon are detected at 2:1 as expected for \sio2. Some carbon from adventitious sources is detected. Neither fluorine nor aluminum are present above the detection limit ($\approx 0.2$\%), and sulfur is not shown in any plots as it was not found in any sample. Note that the silicon peak consists of two subpeaks; the larger peak at $\sim$104 eV corresponds to \sio2, while the smaller peak at 102 eV could be attributed to surface contamination from either silicon (oxy)nitride or organic bonds formed at the end of thermal oxidation \cite{NIST_XPS}.


After ALE, the O:Si ratio is unchanged at 2:1, and the overall quantity of these elements decreases by 1.6\%. The carbon concentration in the ALE-treated film is the same as in the untreated film to within uncertainty. Fluorine and aluminum are found to compose less than $2\%$ of the sample surface. The Si sub-peak is not observed, consistent with it originating from a surface contaminant. An oxygen sub-peak appears at 531.5 eV, comprising 11\% of the total peak area, which we assign to aluminum species \cite{NIST_XPS}.

\begin{table}
    \newcolumntype{Y}{>{\centering\arraybackslash}X}
    \begin{tabularx}{14cm}{>{\centering} YYYYYYY}
        \hline\hline
        \multicolumn{2}{c}{Sample} & O (\%) & Si (\%) & C (\%) & F (\%) & Al (\%) \\
        \hline
        \multicolumn{2}{c}{Untreated} & $62.4\pm0.8$ & $31.2\pm0.5$ & $6.4\pm1.4$ & $<0.2$ & $<0.2$ \\
        \multicolumn{2}{c}{ALE} & $61.5\pm0.8$ & $30.5\pm0.5$ & $6.2\pm1.8$ & $1.0\pm0.8$ & $0.7\pm1.2$ \\
        \multicolumn{2}{c}{ALE and O\textsubscript{2} plasma} & $62.6\pm0.7$ & $30.8\pm0.6$ & $5.3\pm1.7$ & $0.40\pm0.5$ & $0.96\pm1.1$ \\
        \hline\hline
    \end{tabularx}
    \caption{XPS surface composition of untreated \sio2, ALE treated, and ALE treated with a post-process oxygen plasma step. The estimated detection limit is $0.2\%$.}
    \label{tab:XPS_data}
\end{table}


After oxygen plasma post-treatment, the surface is found to contain decreased concentrations of carbon and fluorine. The oxygen sub-peak is still present in the same location at 13\% of the area. Therefore, we find that the oxygen treatment is effective in decreasing the amount of fluorine while leaving the remainder of the film composition unchanged. Similar ALE works using TMA and HF have reported approximately 8.5\% F \cite{Rahman:2018} and 4\% Al (Refs. \onlinecite{DuMont:2017,Rahman:2018}) after etching. After oxygen plasma treatment, our process results in 89\% less F and Al contamination compared to the similar works.



\subsection{\texorpdfstring{EPC dependence on \sio2 preparation method}{EPC dependence on SiO2 preparation method}}

\begin{figure}
    \centering{
        \includegraphics[width=3.4in, height=3.0in]{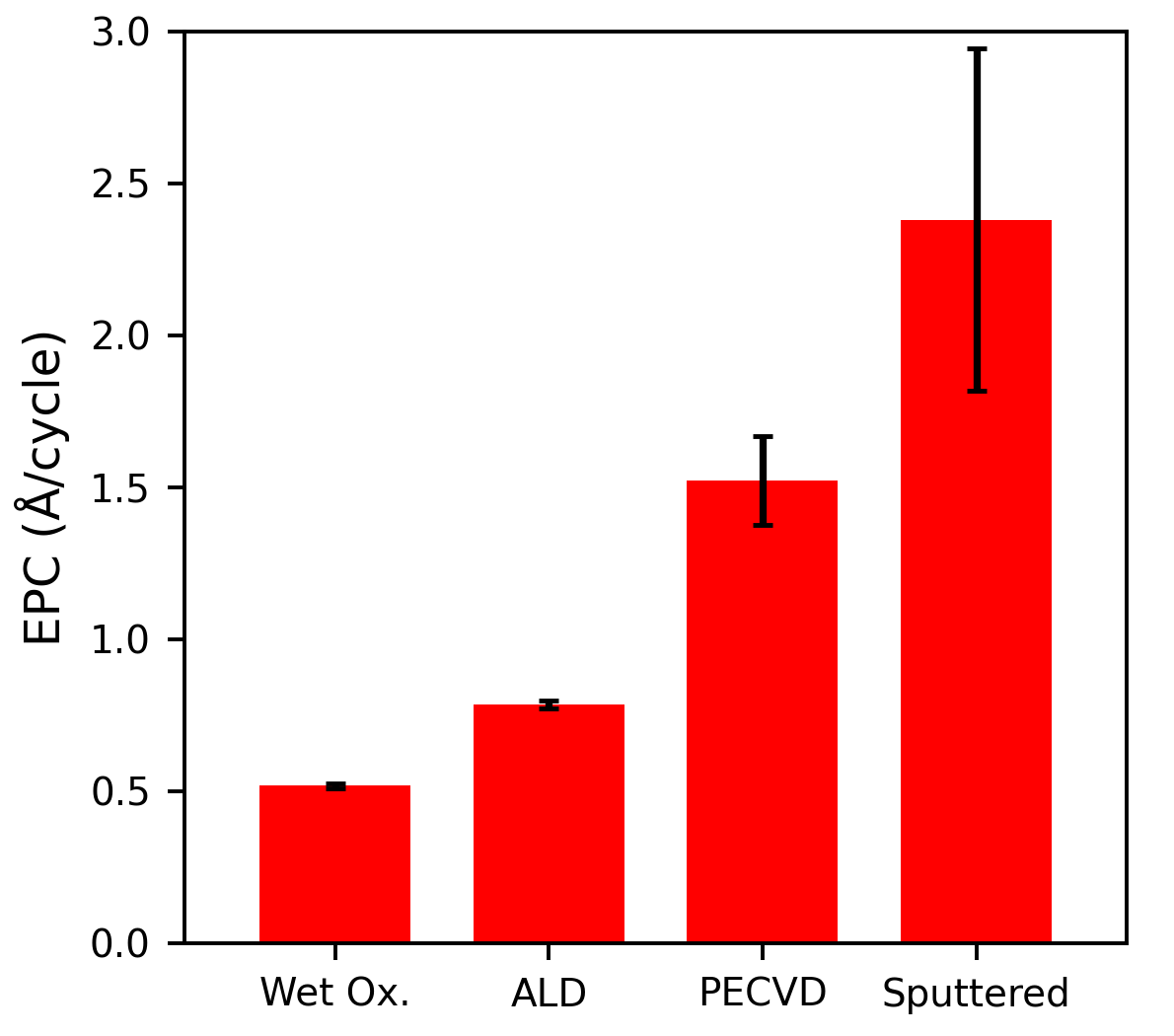}}
    \caption{ALE etch per cycle versus \sio2 preparation method.}
    \label{fig:deposition}
\end{figure}

We measured the EPC on four samples prepared by different methods. It has previously been established that oxides generally exhibit different etch rates during ALE depending on their crystallinity \cite{Murdzek:2020,Murdzek:2021}, and although \sio2 is typically used in amorphous form, different preparation methods may produce different amorphous structures. Our first sample was thermally oxidized as described in \cref{sec:methods} although with water vapor instead of oxygen. The second sample was made using ALD with BDEAS and O\textsubscript{2} plasma at \qty{200}{\degreeCelsius} on the Oxford FlexAL. The third sample was made using PECVD with silane and N\textsubscript{2}O at \qty{200}{\degreeCelsius} in an Oxford Instruments Plasma Technology Plasmalab System 100. The fourth sample was fabricated via sputtering at ambient temperature using an \sio2 target on an AJA Orion ATC 1800 system. The samples were pretreated with 20 cycles of ALE before measurements.

The EPC for each of our four films is shown in \cref{fig:deposition}. We observe that thermally wet-oxidized \sio2 etches the slowest at $0.52\pm0.01$ \AA/cycle. No difference in rate is observed when etching dry-oxidized films. ALD \sio2 has a comparable etch rate of $0.78\pm0.01$ \AA/cycle. PECVD \sio2 exhibits a larger etch rate of $1.52\pm0.15$ \AA/cycle, and sputtered \sio2 has the largest etch rate of $2.38\pm0.56$ \AA/cycle.

A possible explanation for the observed trend is that ALE etch rates are inversely correlated to film density (with lower density films having higher atomic diffusion coefficients), which is in turn affected by the film preparation method. It is known that PECVD \cite{Ceiler:1995} and sputtered \cite{Simurka:2018} \sio2 densities vary based on process conditions and may be less than those of ALD \cite{Dingemans:2011} and thermal films. Because thin-film density is difficult to determine directly, we measured the index of refraction ($n$) of our samples as an indirect measure of density \cite{Taniguchi:1990,Dingemans:2011}, assuming that $n$ is a function of density only. However, no correlation between $n$ and the etch rate was found. Therefore, we are unable to draw a conclusion as to the cause of the varying etch rates. In addition to density, the films may also differ in impurity content (hydrogen \cite{Ceiler:1995,Dingemans:2011} or other decomposition byproducts of silane and BDEAS) and the amorphous structure formed by each preparation process, both of which could affect the etch rate.


\subsection{Surface roughness} \label{sec:roughness}

\begin{figure}
    \centering{
        \phantomsubcaption\label{subfig:afm_a}
        \phantomsubcaption\label{subfig:afm_b}
        \includegraphics[width=5.5in, height=2.5in]{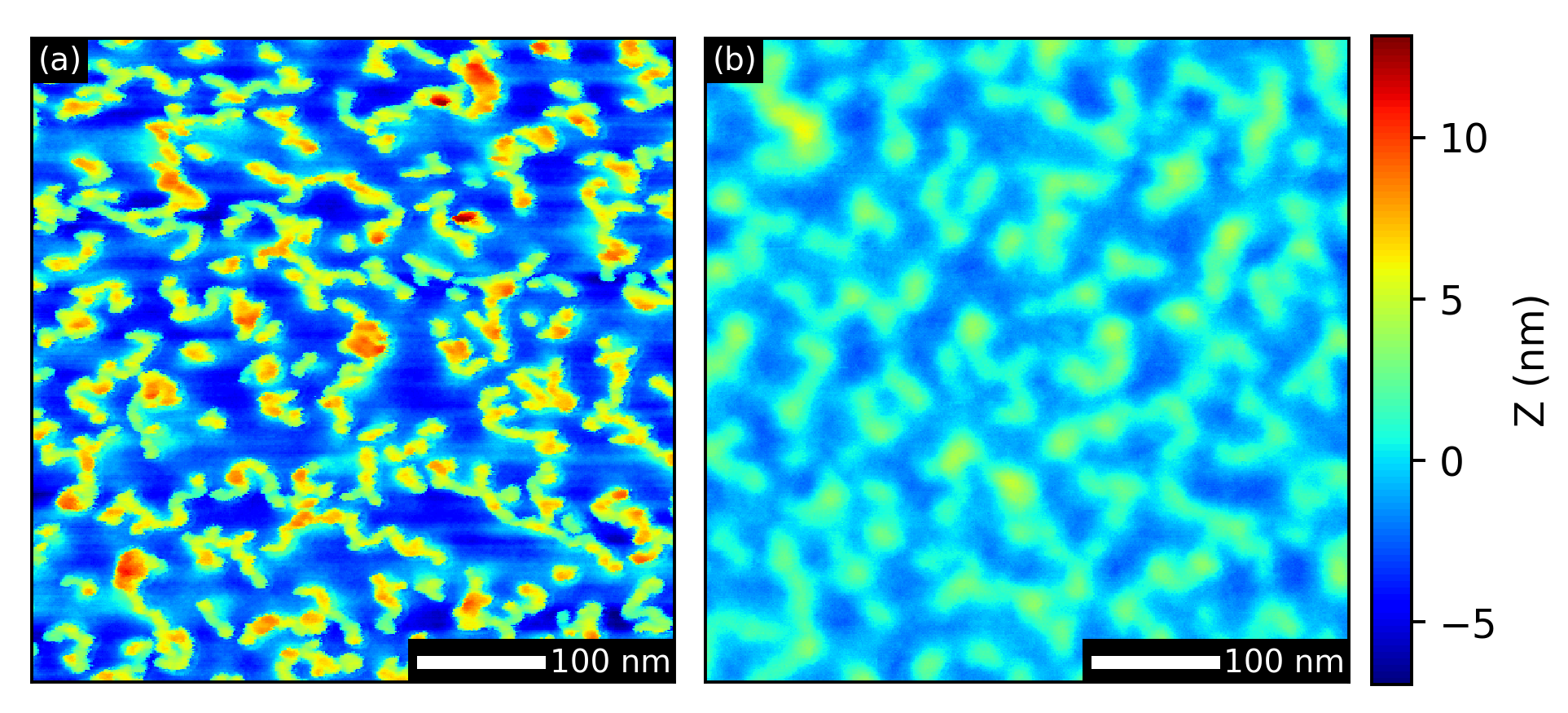}}
    \caption{(a) Surface topography of \sio2 roughened via rapid \sf6 etching, with $R_q$ = 3.49 nm, $R_a$ = 2.95 nm. (b) After treatment with 300 cycles of ALE, corresponding to an etch depth of $12.4 \pm 0.3$ nm. The roughness has decreased, with $R_q$ = 1.33 nm, $R_a$ = 1.08 nm.}
    \label{fig:afm}
\end{figure}

We finally examine the change in surface roughness induced by ALE. We fabricated rough \sio2 films by Si wet-oxidization to a thickness of $227 \pm 1.2$ nm followed by RIE (a 315 second plasma exposure using 40 sccm \sf6 and 150 sccm Ar at 50 W RF table bias) to a thickness of $30.6 \pm 3.9$ nm. The AFM scan of this roughened film is given in \cref{subfig:afm_a}. The surface is an order of magnitude rougher than a pristine \sio2 surface ($R_q=3.49$ nm versus $R_q\lesssim0.30$ nm).

The roughened sample was then treated with 300 cycles of a version of the ALE recipe optimized to minimize the overall process time. This process had the following parameters: \qty{450}{\degreeCelsius} table temperature, 8 s of \textit{in-situ} HF, 15 s of TMA, 750 ms purge times, 2 s pump and gas stabilization times, and 0.5 s for the APC valve to close before TMA dosing. The ICP was left open to minimize particulates generated by the gate valve. This ``rapid'' version of the ALE process has a total cycle time of 29 seconds. After 300 cycles, the \sio2 thickness was $18.2 \pm 3.7$ nm, corresponding to an etch depth of $12.4 \pm 0.3$ nm and EPC of $0.41 \pm 0.01$ \AA/cycle. The post-ALE AFM scan is shown in \cref{subfig:afm_b}. The surface is visibly smoother, with a 62\% decrease in roughness to $R_q=1.33$ nm. Therefore, the isotropic ALE of \sio2 using the recipe reported here was found to exhibit a surface smoothing effect.

\begin{figure}
    \centering{
        \includegraphics[width=3.4in, height=3.0in]{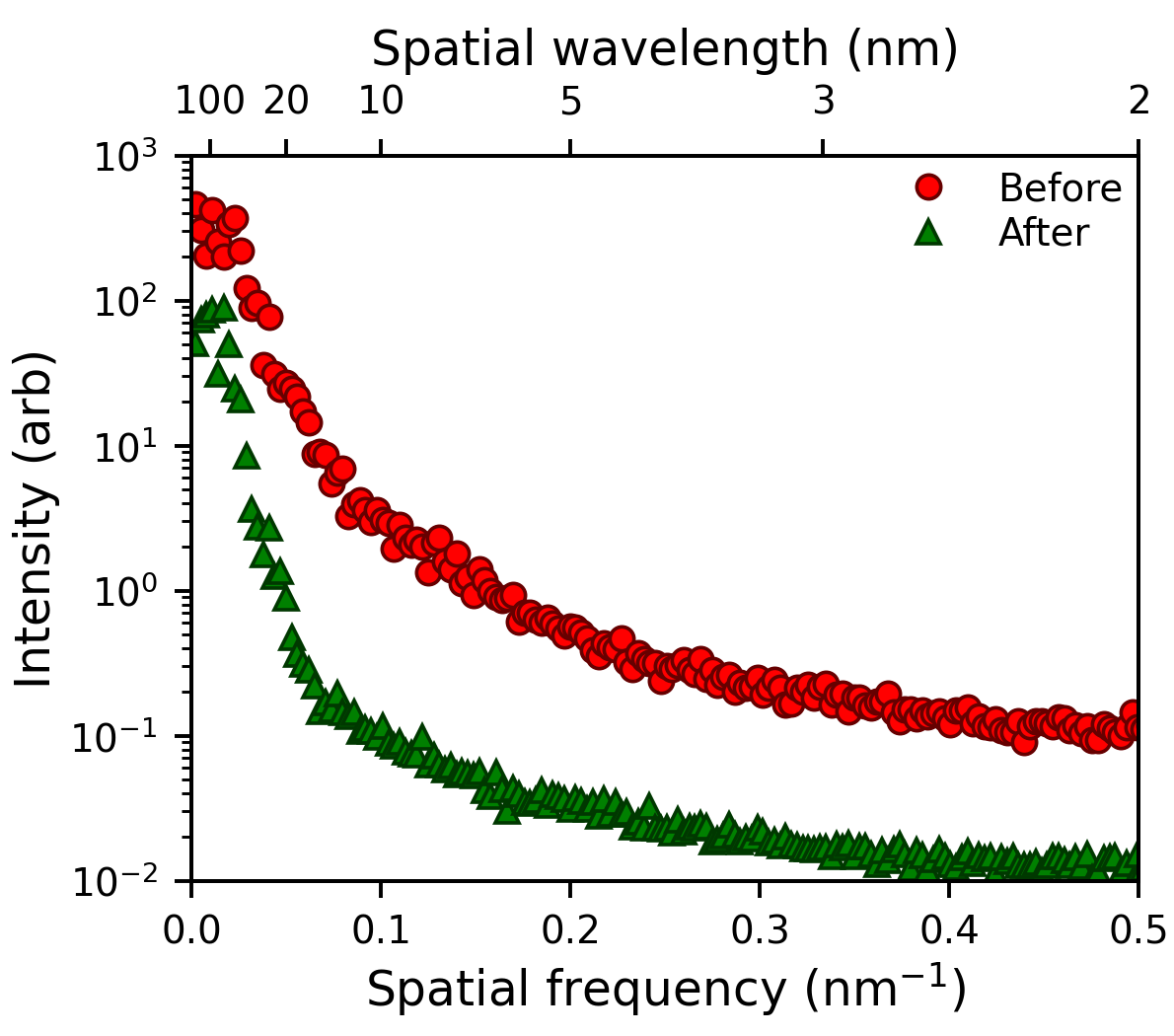}}
    \caption{Power spectral density of surface roughness versus spatial frequency and wavelength for roughened \sio2 surface (red circles) and after 300 cycles of ALE (green triangles).}
    \label{fig:psd}
\end{figure}

To be applied to optical devices, it must also be determined whether the smoothing effect occurs for feature sizes relevant to optical scattering. In \sio2 resonators, the relevant optical wavelengths are generally much larger than the characteristic dimensions of surface features, resulting in a $\sigma^{-2}B^{-1}$ dependence of the roughness-limited $Q$ (Ref. \onlinecite{Gorodetsky:1996}), where $\sigma$ is the root mean square roughness of surface and $B$ is the correlation length of surface features. Therefore, scattering is not only related to the height and quantity of surface features, but also their width, where long period features scatter more strongly than short period ones. 

To determine the degree of smoothing across various feature widths, we calculated the surface power spectral density (PSD) of the roughened \sio2 sample before and after etching the roughened sample. The PSD represents the intensity of surface features versus their spatial wavelength (or frequency) and can therefore be used to determine whether ALE smooths features of various sizes. The PSD was calculated by decomposing four $1000 \times 20$ nm scans taken across the sample into individual line scans, taking the absolute square of the discrete 1D Fourier transform of these scans (with linear fit subtracted) and averaging the results, which are shown in \cref{fig:psd}. We observe a decrease in the PSD intensity across all feature sizes present in the scan. This finding indicates that the ALE process should be suitable for smoothing optical device surfaces regardless of the dominant feature size.

\section{Summary}

In summary, we have reported an isotropic \sio2 ALE recipe compatible with commercial ALD tools based on sequential exposures of Ar/\h2/\sf6 plasma and TMA. The process improves upon prior processes by achieving EPC above 0.5 \AA/cycle while simultaneously eliminating the need for vapor HF, and exhibiting a short cycle time (under 2 minutes) and low TMA pressure requirements (under 250 mTorr). We observe that each process step exhibits saturation and the overall process has a synergy of 100\%. We also examine the etch rate dependence on the preparation method, finding that sputtered \sio2 exhibits the highest etch rate of 2.38 \AA/cycle compared to PECVD, ALD, and thermal \sio2. Additionally, our process results in 89\% less contamination by F and Al than comparable processes. Surface smoothing of a roughened surface is demonstrated.  Lastly, because the process is compatible with commercial ALD hardware, performing the process on the wafer scale is limited only by the maximum wafer size of available ALD tools. With the observation of simultaneously low surface contamination and surface smoothing, this work paves the way for the use of \sio2 ALE as a post-processing step to improve photonic device performance.


\section*{Acknowledgements}

This work was supported by the National Science Foundation through award \#2234390 and by the US Air Force Office of Scientific Research through award \#FA9550-23-1-0625. We gratefully acknowledge the critical support and infrastructure provided for this work by Kavli Nanoscience Institute and the Molecular Materials Research Center of the Beckman Institute at the California Institute of Technology for the use of their facilities.

\bibliography{bib}

\begin{thebibliography}{52}%
\makeatletter
\providecommand \@ifxundefined [1]{%
 \@ifx{#1\undefined}
}%
\providecommand \@ifnum [1]{%
 \ifnum #1\expandafter \@firstoftwo
 \else \expandafter \@secondoftwo
 \fi
}%
\providecommand \@ifx [1]{%
 \ifx #1\expandafter \@firstoftwo
 \else \expandafter \@secondoftwo
 \fi
}%
\providecommand \natexlab [1]{#1}%
\providecommand \enquote  [1]{``#1''}%
\providecommand \bibnamefont  [1]{#1}%
\providecommand \bibfnamefont [1]{#1}%
\providecommand \citenamefont [1]{#1}%
\providecommand \href@noop [0]{\@secondoftwo}%
\providecommand \href [0]{\begingroup \@sanitize@url \@href}%
\providecommand \@href[1]{\@@startlink{#1}\@@href}%
\providecommand \@@href[1]{\endgroup#1\@@endlink}%
\providecommand \@sanitize@url [0]{\catcode `\\12\catcode `\$12\catcode `\&12\catcode `\#12\catcode `\^12\catcode `\_12\catcode `\%12\relax}%
\providecommand \@@startlink[1]{}%
\providecommand \@@endlink[0]{}%
\providecommand \url  [0]{\begingroup\@sanitize@url \@url }%
\providecommand \@url [1]{\endgroup\@href {#1}{\urlprefix }}%
\providecommand \urlprefix  [0]{URL }%
\providecommand \Eprint [0]{\href }%
\providecommand \doibase [0]{https://doi.org/}%
\providecommand \selectlanguage [0]{\@gobble}%
\providecommand \bibinfo  [0]{\@secondoftwo}%
\providecommand \bibfield  [0]{\@secondoftwo}%
\providecommand \translation [1]{[#1]}%
\providecommand \BibitemOpen [0]{}%
\providecommand \bibitemStop [0]{}%
\providecommand \bibitemNoStop [0]{.\EOS\space}%
\providecommand \EOS [0]{\spacefactor3000\relax}%
\providecommand \BibitemShut  [1]{\csname bibitem#1\endcsname}%
\let\auto@bib@innerbib\@empty
\bibitem [{\citenamefont {Oehrlein}\ \emph {et~al.}(2015)\citenamefont {Oehrlein}, \citenamefont {Metzler},\ and\ \citenamefont {Li}}]{Oehrlein:2015}%
  \BibitemOpen
  \bibfield  {author} {\bibinfo {author} {\bibfnamefont {G.~S.}\ \bibnamefont {Oehrlein}}, \bibinfo {author} {\bibfnamefont {D.}~\bibnamefont {Metzler}},\ and\ \bibinfo {author} {\bibfnamefont {C.}~\bibnamefont {Li}},\ }\href {https://doi.org/10.1149/2.0061506jss} {\bibfield  {journal} {\bibinfo  {journal} {ECS Journal of Solid State Science and Technology}\ }\textbf {\bibinfo {volume} {4}},\ \bibinfo {pages} {N5041} (\bibinfo {year} {2015})}\BibitemShut {NoStop}%
\bibitem [{\citenamefont {Fang}\ \emph {et~al.}(2018)\citenamefont {Fang}, \citenamefont {Cao}, \citenamefont {Wu},\ and\ \citenamefont {Li}}]{Fang:2018}%
  \BibitemOpen
  \bibfield  {author} {\bibinfo {author} {\bibfnamefont {C.}~\bibnamefont {Fang}}, \bibinfo {author} {\bibfnamefont {Y.}~\bibnamefont {Cao}}, \bibinfo {author} {\bibfnamefont {D.}~\bibnamefont {Wu}},\ and\ \bibinfo {author} {\bibfnamefont {A.}~\bibnamefont {Li}},\ }\href {https://doi.org/https://doi.org/10.1016/j.pnsc.2018.11.003} {\bibfield  {journal} {\bibinfo  {journal} {Progress in Natural Science: Materials International}\ }\textbf {\bibinfo {volume} {28}},\ \bibinfo {pages} {667} (\bibinfo {year} {2018})}\BibitemShut {NoStop}%
\bibitem [{\citenamefont {George}(2020)}]{George:2020}%
  \BibitemOpen
  \bibfield  {author} {\bibinfo {author} {\bibfnamefont {S.~M.}\ \bibnamefont {George}},\ }\href {https://doi.org/10.1021/acs.accounts.0c00084} {\bibfield  {journal} {\bibinfo  {journal} {Accounts of Chemical Research}\ }\textbf {\bibinfo {volume} {53}},\ \bibinfo {pages} {1151} (\bibinfo {year} {2020})},\ \bibinfo {note} {pMID: 32476413}\BibitemShut {NoStop}%
\bibitem [{\citenamefont {Lill}(2021)}]{Lill:2021}%
  \BibitemOpen
  \bibfield  {author} {\bibinfo {author} {\bibfnamefont {T.}~\bibnamefont {Lill}},\ }\bibinfo {title} {Atomic layer processing: Semiconductor dry etching technology}\ (\bibinfo  {publisher} {Wiley},\ \bibinfo {year} {2021})\ Chap.~\bibinfo {chapter} {4}\BibitemShut {NoStop}%
\bibitem [{\citenamefont {Fischer}\ \emph {et~al.}(2021)\citenamefont {Fischer}, \citenamefont {Routzahn}, \citenamefont {George},\ and\ \citenamefont {Lill}}]{Fischer:2021}%
  \BibitemOpen
  \bibfield  {author} {\bibinfo {author} {\bibfnamefont {A.}~\bibnamefont {Fischer}}, \bibinfo {author} {\bibfnamefont {A.}~\bibnamefont {Routzahn}}, \bibinfo {author} {\bibfnamefont {S.~M.}\ \bibnamefont {George}},\ and\ \bibinfo {author} {\bibfnamefont {T.}~\bibnamefont {Lill}},\ }\href {https://doi.org/10.1116/6.0000894} {\bibfield  {journal} {\bibinfo  {journal} {Journal of Vacuum Science \& Technology A}\ }\textbf {\bibinfo {volume} {39}},\ \bibinfo {pages} {030801} (\bibinfo {year} {2021})}\BibitemShut {NoStop}%
\bibitem [{\citenamefont {Fischer}\ and\ \citenamefont {Lill}(2023)}]{Fischer:2023}%
  \BibitemOpen
  \bibfield  {author} {\bibinfo {author} {\bibfnamefont {A.}~\bibnamefont {Fischer}}\ and\ \bibinfo {author} {\bibfnamefont {T.}~\bibnamefont {Lill}},\ }\href {https://doi.org/10.1063/5.0158785} {\bibfield  {journal} {\bibinfo  {journal} {Physics of Plasmas}\ }\textbf {\bibinfo {volume} {30}},\ \bibinfo {pages} {080601} (\bibinfo {year} {2023})}\BibitemShut {NoStop}%
\bibitem [{\citenamefont {Kanarik}\ \emph {et~al.}(2018)\citenamefont {Kanarik}, \citenamefont {Tan},\ and\ \citenamefont {Gottscho}}]{Kanarik:2018}%
  \BibitemOpen
  \bibfield  {author} {\bibinfo {author} {\bibfnamefont {K.~J.}\ \bibnamefont {Kanarik}}, \bibinfo {author} {\bibfnamefont {S.}~\bibnamefont {Tan}},\ and\ \bibinfo {author} {\bibfnamefont {R.~A.}\ \bibnamefont {Gottscho}},\ }\href {https://doi.org/10.1021/acs.jpclett.8b00997} {\bibfield  {journal} {\bibinfo  {journal} {The Journal of Physical Chemistry Letters}\ }\textbf {\bibinfo {volume} {9}},\ \bibinfo {pages} {4814} (\bibinfo {year} {2018})},\ \bibinfo {note} {pMID: 30095919}\BibitemShut {NoStop}%
\bibitem [{\citenamefont {Gerritsen}\ \emph {et~al.}(2022)\citenamefont {Gerritsen}, \citenamefont {Chittock}, \citenamefont {Vandalon}, \citenamefont {Verheijen}, \citenamefont {Knoops}, \citenamefont {Kessels},\ and\ \citenamefont {Mackus}}]{Gerritsen:2022}%
  \BibitemOpen
  \bibfield  {author} {\bibinfo {author} {\bibfnamefont {S.~H.}\ \bibnamefont {Gerritsen}}, \bibinfo {author} {\bibfnamefont {N.~J.}\ \bibnamefont {Chittock}}, \bibinfo {author} {\bibfnamefont {V.}~\bibnamefont {Vandalon}}, \bibinfo {author} {\bibfnamefont {M.~A.}\ \bibnamefont {Verheijen}}, \bibinfo {author} {\bibfnamefont {H.~C.~M.}\ \bibnamefont {Knoops}}, \bibinfo {author} {\bibfnamefont {W.~M.~M.}\ \bibnamefont {Kessels}},\ and\ \bibinfo {author} {\bibfnamefont {A.~J.~M.}\ \bibnamefont {Mackus}},\ }\href {https://doi.org/10.1021/acsanm.2c04025} {\bibfield  {journal} {\bibinfo  {journal} {ACS Applied Nano Materials}\ }\textbf {\bibinfo {volume} {5}},\ \bibinfo {pages} {18116} (\bibinfo {year} {2022})}\BibitemShut {NoStop}%
\bibitem [{\citenamefont {Chen}\ \emph {et~al.}(2022)\citenamefont {Chen}, \citenamefont {Cho}, \citenamefont {Chang}, \citenamefont {Su}, \citenamefont {Chu}, \citenamefont {Kei},\ and\ \citenamefont {Li}}]{Chen:2022}%
  \BibitemOpen
  \bibfield  {author} {\bibinfo {author} {\bibfnamefont {C.-W.}\ \bibnamefont {Chen}}, \bibinfo {author} {\bibfnamefont {W.-H.}\ \bibnamefont {Cho}}, \bibinfo {author} {\bibfnamefont {C.-Y.}\ \bibnamefont {Chang}}, \bibinfo {author} {\bibfnamefont {C.-Y.}\ \bibnamefont {Su}}, \bibinfo {author} {\bibfnamefont {N.-N.}\ \bibnamefont {Chu}}, \bibinfo {author} {\bibfnamefont {C.-C.}\ \bibnamefont {Kei}},\ and\ \bibinfo {author} {\bibfnamefont {B.-R.}\ \bibnamefont {Li}},\ }\href {https://doi.org/10.1116/6.0002210} {\bibfield  {journal} {\bibinfo  {journal} {Journal of Vacuum Science \& Technology A}\ }\textbf {\bibinfo {volume} {41}},\ \bibinfo {pages} {012602} (\bibinfo {year} {2022})}\BibitemShut {NoStop}%
\bibitem [{\citenamefont {Pacco}\ \emph {et~al.}(2019)\citenamefont {Pacco}, \citenamefont {Akanishi}, \citenamefont {Le}, \citenamefont {Kesters}, \citenamefont {Murdoch},\ and\ \citenamefont {Holsteyns}}]{Pacco:2019}%
  \BibitemOpen
  \bibfield  {author} {\bibinfo {author} {\bibfnamefont {A.}~\bibnamefont {Pacco}}, \bibinfo {author} {\bibfnamefont {Y.}~\bibnamefont {Akanishi}}, \bibinfo {author} {\bibfnamefont {Q.~T.}\ \bibnamefont {Le}}, \bibinfo {author} {\bibfnamefont {E.}~\bibnamefont {Kesters}}, \bibinfo {author} {\bibfnamefont {G.}~\bibnamefont {Murdoch}},\ and\ \bibinfo {author} {\bibfnamefont {F.}~\bibnamefont {Holsteyns}},\ }\href {https://doi.org/https://doi.org/10.1016/j.mee.2019.111131} {\bibfield  {journal} {\bibinfo  {journal} {Microelectronic Engineering}\ }\textbf {\bibinfo {volume} {217}},\ \bibinfo {pages} {111131} (\bibinfo {year} {2019})}\BibitemShut {NoStop}%
\bibitem [{\citenamefont {Gong}\ \emph {et~al.}(2018)\citenamefont {Gong}, \citenamefont {Venkatraman},\ and\ \citenamefont {Akolkar}}]{Gong:2018}%
  \BibitemOpen
  \bibfield  {author} {\bibinfo {author} {\bibfnamefont {Y.}~\bibnamefont {Gong}}, \bibinfo {author} {\bibfnamefont {K.}~\bibnamefont {Venkatraman}},\ and\ \bibinfo {author} {\bibfnamefont {R.}~\bibnamefont {Akolkar}},\ }\href {https://doi.org/10.1149/2.0901807jes} {\bibfield  {journal} {\bibinfo  {journal} {Journal of The Electrochemical Society}\ }\textbf {\bibinfo {volume} {165}},\ \bibinfo {pages} {D282} (\bibinfo {year} {2018})}\BibitemShut {NoStop}%
\bibitem [{\citenamefont {Mohimi}\ \emph {et~al.}(2018)\citenamefont {Mohimi}, \citenamefont {Chu}, \citenamefont {Trinh}, \citenamefont {Babar}, \citenamefont {Girolami},\ and\ \citenamefont {Abelson}}]{Mohimi:2018}%
  \BibitemOpen
  \bibfield  {author} {\bibinfo {author} {\bibfnamefont {E.}~\bibnamefont {Mohimi}}, \bibinfo {author} {\bibfnamefont {X.~I.}\ \bibnamefont {Chu}}, \bibinfo {author} {\bibfnamefont {B.~B.}\ \bibnamefont {Trinh}}, \bibinfo {author} {\bibfnamefont {S.}~\bibnamefont {Babar}}, \bibinfo {author} {\bibfnamefont {G.~S.}\ \bibnamefont {Girolami}},\ and\ \bibinfo {author} {\bibfnamefont {J.~R.}\ \bibnamefont {Abelson}},\ }\href {https://doi.org/10.1149/2.0211809jss} {\bibfield  {journal} {\bibinfo  {journal} {ECS Journal of Solid State Science and Technology}\ }\textbf {\bibinfo {volume} {7}},\ \bibinfo {pages} {P491} (\bibinfo {year} {2018})}\BibitemShut {NoStop}%
\bibitem [{\citenamefont {Sheil}\ \emph {et~al.}(2021)\citenamefont {Sheil}, \citenamefont {Martirez}, \citenamefont {Sang}, \citenamefont {Carter},\ and\ \citenamefont {Chang}}]{Sheil:2021}%
  \BibitemOpen
  \bibfield  {author} {\bibinfo {author} {\bibfnamefont {R.}~\bibnamefont {Sheil}}, \bibinfo {author} {\bibfnamefont {J.~M.~P.}\ \bibnamefont {Martirez}}, \bibinfo {author} {\bibfnamefont {X.}~\bibnamefont {Sang}}, \bibinfo {author} {\bibfnamefont {E.~A.}\ \bibnamefont {Carter}},\ and\ \bibinfo {author} {\bibfnamefont {J.~P.}\ \bibnamefont {Chang}},\ }\href {https://doi.org/10.1021/acs.jpcc.0c08932} {\bibfield  {journal} {\bibinfo  {journal} {The Journal of Physical Chemistry C}\ }\textbf {\bibinfo {volume} {125}},\ \bibinfo {pages} {1819} (\bibinfo {year} {2021})}\BibitemShut {NoStop}%
\bibitem [{\citenamefont {Johnson}\ and\ \citenamefont {George}(2017)}]{Johnson:2017}%
  \BibitemOpen
  \bibfield  {author} {\bibinfo {author} {\bibfnamefont {N.~R.}\ \bibnamefont {Johnson}}\ and\ \bibinfo {author} {\bibfnamefont {S.~M.}\ \bibnamefont {George}},\ }\href {https://doi.org/10.1021/acsami.7b09161} {\bibfield  {journal} {\bibinfo  {journal} {ACS Applied Materials \& Interfaces}\ }\textbf {\bibinfo {volume} {9}},\ \bibinfo {pages} {34435} (\bibinfo {year} {2017})},\ \bibinfo {note} {pMID: 28876892}\BibitemShut {NoStop}%
\bibitem [{\citenamefont {Xie}\ \emph {et~al.}(2018)\citenamefont {Xie}, \citenamefont {Lemaire},\ and\ \citenamefont {Parsons}}]{Xie:2018}%
  \BibitemOpen
  \bibfield  {author} {\bibinfo {author} {\bibfnamefont {W.}~\bibnamefont {Xie}}, \bibinfo {author} {\bibfnamefont {P.~C.}\ \bibnamefont {Lemaire}},\ and\ \bibinfo {author} {\bibfnamefont {G.~N.}\ \bibnamefont {Parsons}},\ }\href {https://doi.org/10.1021/acsami.7b19024} {\bibfield  {journal} {\bibinfo  {journal} {ACS Applied Materials \& Interfaces}\ }\textbf {\bibinfo {volume} {10}},\ \bibinfo {pages} {9147} (\bibinfo {year} {2018})},\ \bibinfo {note} {pMID: 29461793}\BibitemShut {NoStop}%
\bibitem [{\citenamefont {Athavale}\ and\ \citenamefont {Economou}(1996)}]{Athavale:1996}%
  \BibitemOpen
  \bibfield  {author} {\bibinfo {author} {\bibfnamefont {S.~D.}\ \bibnamefont {Athavale}}\ and\ \bibinfo {author} {\bibfnamefont {D.~J.}\ \bibnamefont {Economou}},\ }\href {https://doi.org/10.1116/1.588651} {\bibfield  {journal} {\bibinfo  {journal} {Journal of Vacuum Science \& Technology B: Microelectronics and Nanometer Structures Processing, Measurement, and Phenomena}\ }\textbf {\bibinfo {volume} {14}},\ \bibinfo {pages} {3702} (\bibinfo {year} {1996})}\BibitemShut {NoStop}%
\bibitem [{\citenamefont {Park}\ \emph {et~al.}(2005)\citenamefont {Park}, \citenamefont {Lee},\ and\ \citenamefont {Yeom}}]{Park:2005}%
  \BibitemOpen
  \bibfield  {author} {\bibinfo {author} {\bibfnamefont {S.~D.}\ \bibnamefont {Park}}, \bibinfo {author} {\bibfnamefont {D.~H.}\ \bibnamefont {Lee}},\ and\ \bibinfo {author} {\bibfnamefont {G.~Y.}\ \bibnamefont {Yeom}},\ }\href {https://doi.org/10.1149/1.1938848} {\bibfield  {journal} {\bibinfo  {journal} {Electrochemical and Solid-State Letters}\ }\textbf {\bibinfo {volume} {8}},\ \bibinfo {pages} {C106} (\bibinfo {year} {2005})}\BibitemShut {NoStop}%
\bibitem [{\citenamefont {Park}\ \emph {et~al.}(2006)\citenamefont {Park}, \citenamefont {Oh}, \citenamefont {Bae}, \citenamefont {Yeom}, \citenamefont {Kim}, \citenamefont {Song},\ and\ \citenamefont {Jang}}]{Park:2006}%
  \BibitemOpen
  \bibfield  {author} {\bibinfo {author} {\bibfnamefont {S.~D.}\ \bibnamefont {Park}}, \bibinfo {author} {\bibfnamefont {C.~K.}\ \bibnamefont {Oh}}, \bibinfo {author} {\bibfnamefont {J.~W.}\ \bibnamefont {Bae}}, \bibinfo {author} {\bibfnamefont {G.~Y.}\ \bibnamefont {Yeom}}, \bibinfo {author} {\bibfnamefont {T.~W.}\ \bibnamefont {Kim}}, \bibinfo {author} {\bibfnamefont {J.~I.}\ \bibnamefont {Song}},\ and\ \bibinfo {author} {\bibfnamefont {J.~H.}\ \bibnamefont {Jang}},\ }\href {https://doi.org/10.1063/1.2221504} {\bibfield  {journal} {\bibinfo  {journal} {Applied Physics Letters}\ }\textbf {\bibinfo {volume} {89}},\ \bibinfo {pages} {043109} (\bibinfo {year} {2006})}\BibitemShut {NoStop}%
\bibitem [{\citenamefont {Ko}\ and\ \citenamefont {Pang}(1993)}]{Ko:1993}%
  \BibitemOpen
  \bibfield  {author} {\bibinfo {author} {\bibfnamefont {K.~K.}\ \bibnamefont {Ko}}\ and\ \bibinfo {author} {\bibfnamefont {S.~W.}\ \bibnamefont {Pang}},\ }\href {https://doi.org/10.1116/1.586889} {\bibfield  {journal} {\bibinfo  {journal} {Journal of Vacuum Science \& Technology B: Microelectronics and Nanometer Structures Processing, Measurement, and Phenomena}\ }\textbf {\bibinfo {volume} {11}},\ \bibinfo {pages} {2275} (\bibinfo {year} {1993})}\BibitemShut {NoStop}%
\bibitem [{\citenamefont {Aoyagi}\ \emph {et~al.}(1992)\citenamefont {Aoyagi}, \citenamefont {Shinmura}, \citenamefont {Kawasaki}, \citenamefont {Tanaka}, \citenamefont {Gamo}, \citenamefont {Namba},\ and\ \citenamefont {Nakamoto}}]{Aoyagi:1992}%
  \BibitemOpen
  \bibfield  {author} {\bibinfo {author} {\bibfnamefont {Y.}~\bibnamefont {Aoyagi}}, \bibinfo {author} {\bibfnamefont {K.}~\bibnamefont {Shinmura}}, \bibinfo {author} {\bibfnamefont {K.}~\bibnamefont {Kawasaki}}, \bibinfo {author} {\bibfnamefont {T.}~\bibnamefont {Tanaka}}, \bibinfo {author} {\bibfnamefont {K.}~\bibnamefont {Gamo}}, \bibinfo {author} {\bibfnamefont {S.}~\bibnamefont {Namba}},\ and\ \bibinfo {author} {\bibfnamefont {I.}~\bibnamefont {Nakamoto}},\ }\href {https://doi.org/10.1063/1.106477} {\bibfield  {journal} {\bibinfo  {journal} {Applied Physics Letters}\ }\textbf {\bibinfo {volume} {60}},\ \bibinfo {pages} {968} (\bibinfo {year} {1992})}\BibitemShut {NoStop}%
\bibitem [{\citenamefont {Meguro}\ \emph {et~al.}(1993)\citenamefont {Meguro}, \citenamefont {Ishii}, \citenamefont {Kodama}, \citenamefont {Yamamoto}, \citenamefont {Gamo},\ and\ \citenamefont {Aoyagi}}]{Meguro:1993}%
  \BibitemOpen
  \bibfield  {author} {\bibinfo {author} {\bibfnamefont {T.}~\bibnamefont {Meguro}}, \bibinfo {author} {\bibfnamefont {M.}~\bibnamefont {Ishii}}, \bibinfo {author} {\bibfnamefont {K.}~\bibnamefont {Kodama}}, \bibinfo {author} {\bibfnamefont {Y.}~\bibnamefont {Yamamoto}}, \bibinfo {author} {\bibfnamefont {K.}~\bibnamefont {Gamo}},\ and\ \bibinfo {author} {\bibfnamefont {Y.}~\bibnamefont {Aoyagi}},\ }\href {https://doi.org/https://doi.org/10.1016/0040-6090(93)90142-C} {\bibfield  {journal} {\bibinfo  {journal} {Thin Solid Films}\ }\textbf {\bibinfo {volume} {225}},\ \bibinfo {pages} {136} (\bibinfo {year} {1993})}\BibitemShut {NoStop}%
\bibitem [{\citenamefont {Min}\ \emph {et~al.}(2013)\citenamefont {Min}, \citenamefont {Kang}, \citenamefont {Kim}, \citenamefont {Jhon}, \citenamefont {Jhon},\ and\ \citenamefont {Yeom}}]{Min:2013}%
  \BibitemOpen
  \bibfield  {author} {\bibinfo {author} {\bibfnamefont {K.}~\bibnamefont {Min}}, \bibinfo {author} {\bibfnamefont {S.}~\bibnamefont {Kang}}, \bibinfo {author} {\bibfnamefont {J.}~\bibnamefont {Kim}}, \bibinfo {author} {\bibfnamefont {Y.}~\bibnamefont {Jhon}}, \bibinfo {author} {\bibfnamefont {M.}~\bibnamefont {Jhon}},\ and\ \bibinfo {author} {\bibfnamefont {G.}~\bibnamefont {Yeom}},\ }\href {https://doi.org/https://doi.org/10.1016/j.mee.2013.03.170} {\bibfield  {journal} {\bibinfo  {journal} {Microelectronic Engineering}\ }\textbf {\bibinfo {volume} {110}},\ \bibinfo {pages} {457} (\bibinfo {year} {2013})}\BibitemShut {NoStop}%
\bibitem [{\citenamefont {Lee}\ \emph {et~al.}(2016)\citenamefont {Lee}, \citenamefont {Huffman},\ and\ \citenamefont {George}}]{Lee:2016}%
  \BibitemOpen
  \bibfield  {author} {\bibinfo {author} {\bibfnamefont {Y.}~\bibnamefont {Lee}}, \bibinfo {author} {\bibfnamefont {C.}~\bibnamefont {Huffman}},\ and\ \bibinfo {author} {\bibfnamefont {S.~M.}\ \bibnamefont {George}},\ }\href {https://doi.org/10.1021/acs.chemmater.6b02543} {\bibfield  {journal} {\bibinfo  {journal} {Chemistry of Materials}\ }\textbf {\bibinfo {volume} {28}},\ \bibinfo {pages} {7657} (\bibinfo {year} {2016})}\BibitemShut {NoStop}%
\bibitem [{\citenamefont {Hennessy}\ \emph {et~al.}(2017)\citenamefont {Hennessy}, \citenamefont {Moore}, \citenamefont {Balasubramanian}, \citenamefont {Jewell}, \citenamefont {France},\ and\ \citenamefont {Nikzad}}]{Hennessy:2017}%
  \BibitemOpen
  \bibfield  {author} {\bibinfo {author} {\bibfnamefont {J.}~\bibnamefont {Hennessy}}, \bibinfo {author} {\bibfnamefont {C.~S.}\ \bibnamefont {Moore}}, \bibinfo {author} {\bibfnamefont {K.}~\bibnamefont {Balasubramanian}}, \bibinfo {author} {\bibfnamefont {A.~D.}\ \bibnamefont {Jewell}}, \bibinfo {author} {\bibfnamefont {K.}~\bibnamefont {France}},\ and\ \bibinfo {author} {\bibfnamefont {S.}~\bibnamefont {Nikzad}},\ }\href {https://doi.org/10.1116/1.4986945} {\bibfield  {journal} {\bibinfo  {journal} {Journal of Vacuum Science \& Technology A}\ }\textbf {\bibinfo {volume} {35}},\ \bibinfo {pages} {041512} (\bibinfo {year} {2017})}\BibitemShut {NoStop}%
\bibitem [{\citenamefont {Chittock}\ \emph {et~al.}(2020)\citenamefont {Chittock}, \citenamefont {Vos}, \citenamefont {Faraz}, \citenamefont {Kessels}, \citenamefont {Knoops},\ and\ \citenamefont {Mackus}}]{Chittock:2020}%
  \BibitemOpen
  \bibfield  {author} {\bibinfo {author} {\bibfnamefont {N.~J.}\ \bibnamefont {Chittock}}, \bibinfo {author} {\bibfnamefont {M.~F.~J.}\ \bibnamefont {Vos}}, \bibinfo {author} {\bibfnamefont {T.}~\bibnamefont {Faraz}}, \bibinfo {author} {\bibfnamefont {W.~M. M.~E.}\ \bibnamefont {Kessels}}, \bibinfo {author} {\bibfnamefont {H.~C.~M.}\ \bibnamefont {Knoops}},\ and\ \bibinfo {author} {\bibfnamefont {A.~J.~M.}\ \bibnamefont {Mackus}},\ }\href {https://doi.org/10.1063/5.0022531} {\bibfield  {journal} {\bibinfo  {journal} {Applied Physics Letters}\ }\textbf {\bibinfo {volume} {117}},\ \bibinfo {pages} {162107} (\bibinfo {year} {2020})}\BibitemShut {NoStop}%
\bibitem [{\citenamefont {Metzler}\ \emph {et~al.}(2013)\citenamefont {Metzler}, \citenamefont {Bruce}, \citenamefont {Engelmann}, \citenamefont {Joseph},\ and\ \citenamefont {Oehrlein}}]{Metzler:2013}%
  \BibitemOpen
  \bibfield  {author} {\bibinfo {author} {\bibfnamefont {D.}~\bibnamefont {Metzler}}, \bibinfo {author} {\bibfnamefont {R.~L.}\ \bibnamefont {Bruce}}, \bibinfo {author} {\bibfnamefont {S.}~\bibnamefont {Engelmann}}, \bibinfo {author} {\bibfnamefont {E.~A.}\ \bibnamefont {Joseph}},\ and\ \bibinfo {author} {\bibfnamefont {G.~S.}\ \bibnamefont {Oehrlein}},\ }\href {https://doi.org/10.1116/1.4843575} {\bibfield  {journal} {\bibinfo  {journal} {Journal of Vacuum Science \& Technology A}\ }\textbf {\bibinfo {volume} {32}},\ \bibinfo {pages} {020603} (\bibinfo {year} {2013})}\BibitemShut {NoStop}%
\bibitem [{\citenamefont {Kaler}\ \emph {et~al.}(2017)\citenamefont {Kaler}, \citenamefont {Lou}, \citenamefont {Donnelly},\ and\ \citenamefont {Economou}}]{Kaler:2017}%
  \BibitemOpen
  \bibfield  {author} {\bibinfo {author} {\bibfnamefont {S.~S.}\ \bibnamefont {Kaler}}, \bibinfo {author} {\bibfnamefont {Q.}~\bibnamefont {Lou}}, \bibinfo {author} {\bibfnamefont {V.~M.}\ \bibnamefont {Donnelly}},\ and\ \bibinfo {author} {\bibfnamefont {D.~J.}\ \bibnamefont {Economou}},\ }\href {https://doi.org/10.1088/1361-6463/aa6f40} {\bibfield  {journal} {\bibinfo  {journal} {Journal of Physics D: Applied Physics}\ }\textbf {\bibinfo {volume} {50}},\ \bibinfo {pages} {234001} (\bibinfo {year} {2017})}\BibitemShut {NoStop}%
\bibitem [{\citenamefont {Koh}\ \emph {et~al.}(2017)\citenamefont {Koh}, \citenamefont {Kim}, \citenamefont {Kim},\ and\ \citenamefont {Chae}}]{Koh:2017}%
  \BibitemOpen
  \bibfield  {author} {\bibinfo {author} {\bibfnamefont {K.}~\bibnamefont {Koh}}, \bibinfo {author} {\bibfnamefont {Y.}~\bibnamefont {Kim}}, \bibinfo {author} {\bibfnamefont {C.-K.}\ \bibnamefont {Kim}},\ and\ \bibinfo {author} {\bibfnamefont {H.}~\bibnamefont {Chae}},\ }\href {https://doi.org/10.1116/1.5003417} {\bibfield  {journal} {\bibinfo  {journal} {Journal of Vacuum Science \& Technology A}\ }\textbf {\bibinfo {volume} {36}},\ \bibinfo {pages} {01B106} (\bibinfo {year} {2017})}\BibitemShut {NoStop}%
\bibitem [{\citenamefont {Lin}\ \emph {et~al.}(2020)\citenamefont {Lin}, \citenamefont {Li}, \citenamefont {Engelmann}, \citenamefont {Bruce}, \citenamefont {Joseph}, \citenamefont {Metzler},\ and\ \citenamefont {Oehrlein}}]{Lin:2020}%
  \BibitemOpen
  \bibfield  {author} {\bibinfo {author} {\bibfnamefont {K.-Y.}\ \bibnamefont {Lin}}, \bibinfo {author} {\bibfnamefont {C.}~\bibnamefont {Li}}, \bibinfo {author} {\bibfnamefont {S.}~\bibnamefont {Engelmann}}, \bibinfo {author} {\bibfnamefont {R.~L.}\ \bibnamefont {Bruce}}, \bibinfo {author} {\bibfnamefont {E.~A.}\ \bibnamefont {Joseph}}, \bibinfo {author} {\bibfnamefont {D.}~\bibnamefont {Metzler}},\ and\ \bibinfo {author} {\bibfnamefont {G.~S.}\ \bibnamefont {Oehrlein}},\ }\href {https://doi.org/10.1116/1.5143247} {\bibfield  {journal} {\bibinfo  {journal} {Journal of Vacuum Science \& Technology A}\ }\textbf {\bibinfo {volume} {38}},\ \bibinfo {pages} {032601} (\bibinfo {year} {2020})}\BibitemShut {NoStop}%
\bibitem [{\citenamefont {Hossain}\ \emph {et~al.}(2023)\citenamefont {Hossain}, \citenamefont {Wang}, \citenamefont {Catherall}, \citenamefont {Leung}, \citenamefont {Knoops}, \citenamefont {Renzas},\ and\ \citenamefont {Minnich}}]{Azmain:2023}%
  \BibitemOpen
  \bibfield  {author} {\bibinfo {author} {\bibfnamefont {A.~A.}\ \bibnamefont {Hossain}}, \bibinfo {author} {\bibfnamefont {H.}~\bibnamefont {Wang}}, \bibinfo {author} {\bibfnamefont {D.~S.}\ \bibnamefont {Catherall}}, \bibinfo {author} {\bibfnamefont {M.}~\bibnamefont {Leung}}, \bibinfo {author} {\bibfnamefont {H.~C.~M.}\ \bibnamefont {Knoops}}, \bibinfo {author} {\bibfnamefont {J.~R.}\ \bibnamefont {Renzas}},\ and\ \bibinfo {author} {\bibfnamefont {A.~J.}\ \bibnamefont {Minnich}},\ }\href {https://doi.org/10.1116/6.0002965} {\bibfield  {journal} {\bibinfo  {journal} {Journal of Vacuum Science \& Technology A}\ }\textbf {\bibinfo {volume} {41}},\ \bibinfo {pages} {062601} (\bibinfo {year} {2023})}\BibitemShut {NoStop}%
\bibitem [{\citenamefont {Yi}\ \emph {et~al.}(2015)\citenamefont {Yi}, \citenamefont {Yang}, \citenamefont {Yang}, \citenamefont {Suh},\ and\ \citenamefont {Vahala}}]{Yi:2015}%
  \BibitemOpen
  \bibfield  {author} {\bibinfo {author} {\bibfnamefont {X.}~\bibnamefont {Yi}}, \bibinfo {author} {\bibfnamefont {Q.-F.}\ \bibnamefont {Yang}}, \bibinfo {author} {\bibfnamefont {K.~Y.}\ \bibnamefont {Yang}}, \bibinfo {author} {\bibfnamefont {M.-G.}\ \bibnamefont {Suh}},\ and\ \bibinfo {author} {\bibfnamefont {K.}~\bibnamefont {Vahala}},\ }\href {https://doi.org/10.1364/OPTICA.2.001078} {\bibfield  {journal} {\bibinfo  {journal} {Optica}\ }\textbf {\bibinfo {volume} {2}},\ \bibinfo {pages} {1078} (\bibinfo {year} {2015})}\BibitemShut {NoStop}%
\bibitem [{\citenamefont {Wu}\ \emph {et~al.}(2020)\citenamefont {Wu}, \citenamefont {Wang}, \citenamefont {Yang}, \citenamefont {xin Ji}, \citenamefont {Shen}, \citenamefont {Bao}, \citenamefont {Gao},\ and\ \citenamefont {Vahala}}]{Wu:2020}%
  \BibitemOpen
  \bibfield  {author} {\bibinfo {author} {\bibfnamefont {L.}~\bibnamefont {Wu}}, \bibinfo {author} {\bibfnamefont {H.}~\bibnamefont {Wang}}, \bibinfo {author} {\bibfnamefont {Q.}~\bibnamefont {Yang}}, \bibinfo {author} {\bibfnamefont {Q.}~\bibnamefont {xin Ji}}, \bibinfo {author} {\bibfnamefont {B.}~\bibnamefont {Shen}}, \bibinfo {author} {\bibfnamefont {C.}~\bibnamefont {Bao}}, \bibinfo {author} {\bibfnamefont {M.}~\bibnamefont {Gao}},\ and\ \bibinfo {author} {\bibfnamefont {K.}~\bibnamefont {Vahala}},\ }\href {https://doi.org/10.1364/OL.394940} {\bibfield  {journal} {\bibinfo  {journal} {Opt. Lett.}\ }\textbf {\bibinfo {volume} {45}},\ \bibinfo {pages} {5129} (\bibinfo {year} {2020})}\BibitemShut {NoStop}%
\bibitem [{\citenamefont {Wu}\ \emph {et~al.}(2023)\citenamefont {Wu}, \citenamefont {Gao}, \citenamefont {Liu}, \citenamefont {Chen}, \citenamefont {Colburn}, \citenamefont {Blauvelt},\ and\ \citenamefont {Vahala}}]{Wu:2023}%
  \BibitemOpen
  \bibfield  {author} {\bibinfo {author} {\bibfnamefont {L.}~\bibnamefont {Wu}}, \bibinfo {author} {\bibfnamefont {M.}~\bibnamefont {Gao}}, \bibinfo {author} {\bibfnamefont {J.-Y.}\ \bibnamefont {Liu}}, \bibinfo {author} {\bibfnamefont {H.-J.}\ \bibnamefont {Chen}}, \bibinfo {author} {\bibfnamefont {K.}~\bibnamefont {Colburn}}, \bibinfo {author} {\bibfnamefont {H.~A.}\ \bibnamefont {Blauvelt}},\ and\ \bibinfo {author} {\bibfnamefont {K.~J.}\ \bibnamefont {Vahala}},\ }\href {https://doi.org/10.1364/OL.492067} {\bibfield  {journal} {\bibinfo  {journal} {Opt. Lett.}\ }\textbf {\bibinfo {volume} {48}},\ \bibinfo {pages} {3511} (\bibinfo {year} {2023})}\BibitemShut {NoStop}%
\bibitem [{\citenamefont {Lee}\ \emph {et~al.}(2012)\citenamefont {Lee}, \citenamefont {Chen}, \citenamefont {Li}, \citenamefont {Yang}, \citenamefont {Jeon}, \citenamefont {Painter},\ and\ \citenamefont {Vahala}}]{Lee:2012}%
  \BibitemOpen
  \bibfield  {author} {\bibinfo {author} {\bibfnamefont {H.}~\bibnamefont {Lee}}, \bibinfo {author} {\bibfnamefont {T.}~\bibnamefont {Chen}}, \bibinfo {author} {\bibfnamefont {J.}~\bibnamefont {Li}}, \bibinfo {author} {\bibfnamefont {K.~Y.}\ \bibnamefont {Yang}}, \bibinfo {author} {\bibfnamefont {S.}~\bibnamefont {Jeon}}, \bibinfo {author} {\bibfnamefont {O.}~\bibnamefont {Painter}},\ and\ \bibinfo {author} {\bibfnamefont {K.~J.}\ \bibnamefont {Vahala}},\ }\href {https://doi.org/10.1038/nphoton.2012.109} {\bibfield  {journal} {\bibinfo  {journal} {Nature Photonics}\ }\textbf {\bibinfo {volume} {6}},\ \bibinfo {pages} {369} (\bibinfo {year} {2012})}\BibitemShut {NoStop}%
\bibitem [{\citenamefont {Cho}\ \emph {et~al.}(2020)\citenamefont {Cho}, \citenamefont {Kim}, \citenamefont {Kim},\ and\ \citenamefont {Chae}}]{Cho:2020}%
  \BibitemOpen
  \bibfield  {author} {\bibinfo {author} {\bibfnamefont {Y.}~\bibnamefont {Cho}}, \bibinfo {author} {\bibfnamefont {Y.}~\bibnamefont {Kim}}, \bibinfo {author} {\bibfnamefont {S.}~\bibnamefont {Kim}},\ and\ \bibinfo {author} {\bibfnamefont {H.}~\bibnamefont {Chae}},\ }\href {https://doi.org/10.1116/1.5132986} {\bibfield  {journal} {\bibinfo  {journal} {Journal of Vacuum Science \& Technology A}\ }\textbf {\bibinfo {volume} {38}},\ \bibinfo {pages} {022604} (\bibinfo {year} {2020})}\BibitemShut {NoStop}%
\bibitem [{\citenamefont {Miyoshi}\ \emph {et~al.}(2021)\citenamefont {Miyoshi}, \citenamefont {Kobayashi}, \citenamefont {Shinoda}, \citenamefont {Kurihara}, \citenamefont {Kawamura}, \citenamefont {Kouzuma},\ and\ \citenamefont {Izawa}}]{Miyoshi:2021}%
  \BibitemOpen
  \bibfield  {author} {\bibinfo {author} {\bibfnamefont {N.}~\bibnamefont {Miyoshi}}, \bibinfo {author} {\bibfnamefont {H.}~\bibnamefont {Kobayashi}}, \bibinfo {author} {\bibfnamefont {K.}~\bibnamefont {Shinoda}}, \bibinfo {author} {\bibfnamefont {M.}~\bibnamefont {Kurihara}}, \bibinfo {author} {\bibfnamefont {K.}~\bibnamefont {Kawamura}}, \bibinfo {author} {\bibfnamefont {Y.}~\bibnamefont {Kouzuma}},\ and\ \bibinfo {author} {\bibfnamefont {M.}~\bibnamefont {Izawa}},\ }\href {https://doi.org/10.1116/6.0001517} {\bibfield  {journal} {\bibinfo  {journal} {Journal of Vacuum Science \& Technology A}\ }\textbf {\bibinfo {volume} {40}},\ \bibinfo {pages} {012601} (\bibinfo {year} {2021})}\BibitemShut {NoStop}%
\bibitem [{\citenamefont {Ohtake}\ \emph {et~al.}(2023)\citenamefont {Ohtake}, \citenamefont {Miyoshi}, \citenamefont {Shinoda}, \citenamefont {Fujisaki},\ and\ \citenamefont {Yamaguchi}}]{Ohtake:2023}%
  \BibitemOpen
  \bibfield  {author} {\bibinfo {author} {\bibfnamefont {H.}~\bibnamefont {Ohtake}}, \bibinfo {author} {\bibfnamefont {N.}~\bibnamefont {Miyoshi}}, \bibinfo {author} {\bibfnamefont {K.}~\bibnamefont {Shinoda}}, \bibinfo {author} {\bibfnamefont {S.}~\bibnamefont {Fujisaki}},\ and\ \bibinfo {author} {\bibfnamefont {Y.}~\bibnamefont {Yamaguchi}},\ }\href {https://doi.org/10.35848/1347-4065/acaed0} {\bibfield  {journal} {\bibinfo  {journal} {Japanese Journal of Applied Physics}\ }\textbf {\bibinfo {volume} {62}},\ \bibinfo {pages} {SG0801} (\bibinfo {year} {2023})}\BibitemShut {NoStop}%
\bibitem [{\citenamefont {Gill}\ \emph {et~al.}(2021)\citenamefont {Gill}, \citenamefont {Kim}, \citenamefont {Gil}, \citenamefont {Kim}, \citenamefont {Jang}, \citenamefont {Kim},\ and\ \citenamefont {Yeom}}]{Gill:2021}%
  \BibitemOpen
  \bibfield  {author} {\bibinfo {author} {\bibfnamefont {Y.~J.}\ \bibnamefont {Gill}}, \bibinfo {author} {\bibfnamefont {D.~S.}\ \bibnamefont {Kim}}, \bibinfo {author} {\bibfnamefont {H.~S.}\ \bibnamefont {Gil}}, \bibinfo {author} {\bibfnamefont {K.~H.}\ \bibnamefont {Kim}}, \bibinfo {author} {\bibfnamefont {Y.~J.}\ \bibnamefont {Jang}}, \bibinfo {author} {\bibfnamefont {Y.~E.}\ \bibnamefont {Kim}},\ and\ \bibinfo {author} {\bibfnamefont {G.~Y.}\ \bibnamefont {Yeom}},\ }\href {https://doi.org/https://doi.org/10.1002/ppap.202100063} {\bibfield  {journal} {\bibinfo  {journal} {Plasma Processes and Polymers}\ }\textbf {\bibinfo {volume} {18}},\ \bibinfo {pages} {2100063} (\bibinfo {year} {2021})}\BibitemShut {NoStop}%
\bibitem [{\citenamefont {Gil}\ \emph {et~al.}(2023)\citenamefont {Gil}, \citenamefont {Kim}, \citenamefont {Jang}, \citenamefont {Kim}, \citenamefont {Kwon}, \citenamefont {Kim}, \citenamefont {Kim},\ and\ \citenamefont {Yeom}}]{Gil:2023}%
  \BibitemOpen
  \bibfield  {author} {\bibinfo {author} {\bibfnamefont {H.~S.}\ \bibnamefont {Gil}}, \bibinfo {author} {\bibfnamefont {D.~S.}\ \bibnamefont {Kim}}, \bibinfo {author} {\bibfnamefont {Y.~J.}\ \bibnamefont {Jang}}, \bibinfo {author} {\bibfnamefont {D.~W.}\ \bibnamefont {Kim}}, \bibinfo {author} {\bibfnamefont {H.~I.}\ \bibnamefont {Kwon}}, \bibinfo {author} {\bibfnamefont {G.~C.}\ \bibnamefont {Kim}}, \bibinfo {author} {\bibfnamefont {D.~W.}\ \bibnamefont {Kim}},\ and\ \bibinfo {author} {\bibfnamefont {G.~Y.}\ \bibnamefont {Yeom}},\ }\href {https://doi.org/10.1038/s41598-023-38359-4} {\bibfield  {journal} {\bibinfo  {journal} {Scientific Reports}\ }\textbf {\bibinfo {volume} {13}},\ \bibinfo {pages} {11599} (\bibinfo {year} {2023})}\BibitemShut {NoStop}%
\bibitem [{\citenamefont {DuMont}\ \emph {et~al.}(2017)\citenamefont {DuMont}, \citenamefont {Marquardt}, \citenamefont {Cano},\ and\ \citenamefont {George}}]{DuMont:2017}%
  \BibitemOpen
  \bibfield  {author} {\bibinfo {author} {\bibfnamefont {J.~W.}\ \bibnamefont {DuMont}}, \bibinfo {author} {\bibfnamefont {A.~E.}\ \bibnamefont {Marquardt}}, \bibinfo {author} {\bibfnamefont {A.~M.}\ \bibnamefont {Cano}},\ and\ \bibinfo {author} {\bibfnamefont {S.~M.}\ \bibnamefont {George}},\ }\href {https://doi.org/10.1021/acsami.7b01259} {\bibfield  {journal} {\bibinfo  {journal} {ACS Applied Materials \& Interfaces}\ }\textbf {\bibinfo {volume} {9}},\ \bibinfo {pages} {10296} (\bibinfo {year} {2017})},\ \bibinfo {note} {pMID: 28240864}\BibitemShut {NoStop}%
\bibitem [{\citenamefont {Rahman}\ \emph {et~al.}(2018)\citenamefont {Rahman}, \citenamefont {Mattson}, \citenamefont {Klesko}, \citenamefont {Dangerfield}, \citenamefont {Rivillon-Amy}, \citenamefont {Smith}, \citenamefont {Hausmann},\ and\ \citenamefont {Chabal}}]{Rahman:2018}%
  \BibitemOpen
  \bibfield  {author} {\bibinfo {author} {\bibfnamefont {R.}~\bibnamefont {Rahman}}, \bibinfo {author} {\bibfnamefont {E.~C.}\ \bibnamefont {Mattson}}, \bibinfo {author} {\bibfnamefont {J.~P.}\ \bibnamefont {Klesko}}, \bibinfo {author} {\bibfnamefont {A.}~\bibnamefont {Dangerfield}}, \bibinfo {author} {\bibfnamefont {S.}~\bibnamefont {Rivillon-Amy}}, \bibinfo {author} {\bibfnamefont {D.~C.}\ \bibnamefont {Smith}}, \bibinfo {author} {\bibfnamefont {D.}~\bibnamefont {Hausmann}},\ and\ \bibinfo {author} {\bibfnamefont {Y.~J.}\ \bibnamefont {Chabal}},\ }\href {https://doi.org/10.1021/acsami.8b10899} {\bibfield  {journal} {\bibinfo  {journal} {ACS Applied Materials \& Interfaces}\ }\textbf {\bibinfo {volume} {10}},\ \bibinfo {pages} {31784} (\bibinfo {year} {2018})},\ \bibinfo {note} {pMID: 30179460}\BibitemShut {NoStop}%
\bibitem [{\citenamefont {Volynets}\ \emph {et~al.}(2020)\citenamefont {Volynets}, \citenamefont {Barsukov}, \citenamefont {Kim}, \citenamefont {Jung}, \citenamefont {Nam}, \citenamefont {Han}, \citenamefont {Huang},\ and\ \citenamefont {Kushner}}]{Volynets:2020}%
  \BibitemOpen
  \bibfield  {author} {\bibinfo {author} {\bibfnamefont {V.}~\bibnamefont {Volynets}}, \bibinfo {author} {\bibfnamefont {Y.}~\bibnamefont {Barsukov}}, \bibinfo {author} {\bibfnamefont {G.}~\bibnamefont {Kim}}, \bibinfo {author} {\bibfnamefont {J.-E.}\ \bibnamefont {Jung}}, \bibinfo {author} {\bibfnamefont {S.~K.}\ \bibnamefont {Nam}}, \bibinfo {author} {\bibfnamefont {K.}~\bibnamefont {Han}}, \bibinfo {author} {\bibfnamefont {S.}~\bibnamefont {Huang}},\ and\ \bibinfo {author} {\bibfnamefont {M.~J.}\ \bibnamefont {Kushner}},\ }\href {https://doi.org/10.1116/1.5125568} {\bibfield  {journal} {\bibinfo  {journal} {Journal of Vacuum Science \& Technology A}\ }\textbf {\bibinfo {volume} {38}},\ \bibinfo {pages} {023007} (\bibinfo {year} {2020})}\BibitemShut {NoStop}%
\bibitem [{\citenamefont {Pankratiev}\ \emph {et~al.}(2020)\citenamefont {Pankratiev}, \citenamefont {Barsukov}, \citenamefont {Kobelev}, \citenamefont {Vinogradov}, \citenamefont {Miroshnikov},\ and\ \citenamefont {Smirnov}}]{Pankratiev:2020}%
  \BibitemOpen
  \bibfield  {author} {\bibinfo {author} {\bibfnamefont {P.~A.}\ \bibnamefont {Pankratiev}}, \bibinfo {author} {\bibfnamefont {Y.~V.}\ \bibnamefont {Barsukov}}, \bibinfo {author} {\bibfnamefont {A.~A.}\ \bibnamefont {Kobelev}}, \bibinfo {author} {\bibfnamefont {A.~Y.}\ \bibnamefont {Vinogradov}}, \bibinfo {author} {\bibfnamefont {I.~V.}\ \bibnamefont {Miroshnikov}},\ and\ \bibinfo {author} {\bibfnamefont {A.~S.}\ \bibnamefont {Smirnov}},\ }\href {https://doi.org/10.1088/1742-6596/1697/1/012222} {\bibfield  {journal} {\bibinfo  {journal} {Journal of Physics: Conference Series}\ }\textbf {\bibinfo {volume} {1697}},\ \bibinfo {pages} {012222} (\bibinfo {year} {2020})}\BibitemShut {NoStop}%
\bibitem [{\citenamefont {Herzinger}\ \emph {et~al.}(1998)\citenamefont {Herzinger}, \citenamefont {Johs}, \citenamefont {McGahan}, \citenamefont {Woollam},\ and\ \citenamefont {Paulson}}]{Herzinger:1998}%
  \BibitemOpen
  \bibfield  {author} {\bibinfo {author} {\bibfnamefont {C.~M.}\ \bibnamefont {Herzinger}}, \bibinfo {author} {\bibfnamefont {B.}~\bibnamefont {Johs}}, \bibinfo {author} {\bibfnamefont {W.~A.}\ \bibnamefont {McGahan}}, \bibinfo {author} {\bibfnamefont {J.~A.}\ \bibnamefont {Woollam}},\ and\ \bibinfo {author} {\bibfnamefont {W.}~\bibnamefont {Paulson}},\ }\href {https://doi.org/10.1063/1.367101} {\bibfield  {journal} {\bibinfo  {journal} {Journal of Applied Physics}\ }\textbf {\bibinfo {volume} {83}},\ \bibinfo {pages} {3323} (\bibinfo {year} {1998})}\BibitemShut {NoStop}%
\bibitem [{\citenamefont {of~Standards}\ and\ \citenamefont {Technology}(2000)}]{NIST_XPS}%
  \BibitemOpen
  \bibfield  {author} {\bibinfo {author} {\bibfnamefont {N.~I.}\ \bibnamefont {of~Standards}}\ and\ \bibinfo {author} {\bibnamefont {Technology}},\ }\href {https://doi.org/10.18434/T4T88K} {\bibinfo {title} {Nist x-ray photoelectron spectroscopy database}} (\bibinfo {year} {2000}),\ \bibinfo {note} {accessed: 2024-04-04}\BibitemShut {NoStop}%
\bibitem [{\citenamefont {Murdzek}\ and\ \citenamefont {George}(2020)}]{Murdzek:2020}%
  \BibitemOpen
  \bibfield  {author} {\bibinfo {author} {\bibfnamefont {J.~A.}\ \bibnamefont {Murdzek}}\ and\ \bibinfo {author} {\bibfnamefont {S.~M.}\ \bibnamefont {George}},\ }\href {https://doi.org/10.1116/1.5135317} {\bibfield  {journal} {\bibinfo  {journal} {Journal of Vacuum Science \& Technology A}\ }\textbf {\bibinfo {volume} {38}},\ \bibinfo {pages} {022608} (\bibinfo {year} {2020})}\BibitemShut {NoStop}%
\bibitem [{\citenamefont {Murdzek}\ \emph {et~al.}(2021)\citenamefont {Murdzek}, \citenamefont {Rajashekhar}, \citenamefont {Makala},\ and\ \citenamefont {George}}]{Murdzek:2021}%
  \BibitemOpen
  \bibfield  {author} {\bibinfo {author} {\bibfnamefont {J.~A.}\ \bibnamefont {Murdzek}}, \bibinfo {author} {\bibfnamefont {A.}~\bibnamefont {Rajashekhar}}, \bibinfo {author} {\bibfnamefont {R.~S.}\ \bibnamefont {Makala}},\ and\ \bibinfo {author} {\bibfnamefont {S.~M.}\ \bibnamefont {George}},\ }\href {https://doi.org/10.1116/6.0000995} {\bibfield  {journal} {\bibinfo  {journal} {Journal of Vacuum Science \& Technology A}\ }\textbf {\bibinfo {volume} {39}},\ \bibinfo {pages} {042602} (\bibinfo {year} {2021})}\BibitemShut {NoStop}%
\bibitem [{\citenamefont {Ceiler}\ \emph {et~al.}(1995)\citenamefont {Ceiler}, \citenamefont {Kohl},\ and\ \citenamefont {Bidstrup}}]{Ceiler:1995}%
  \BibitemOpen
  \bibfield  {author} {\bibinfo {author} {\bibfnamefont {M.~F.}\ \bibnamefont {Ceiler}}, \bibinfo {author} {\bibfnamefont {P.~A.}\ \bibnamefont {Kohl}},\ and\ \bibinfo {author} {\bibfnamefont {S.~A.}\ \bibnamefont {Bidstrup}},\ }\href {https://doi.org/10.1149/1.2044242} {\bibfield  {journal} {\bibinfo  {journal} {Journal of The Electrochemical Society}\ }\textbf {\bibinfo {volume} {142}},\ \bibinfo {pages} {2067} (\bibinfo {year} {1995})}\BibitemShut {NoStop}%
\bibitem [{\citenamefont {{\v{S}}imurka}\ \emph {et~al.}(2018)\citenamefont {{\v{S}}imurka}, \citenamefont {{\v{C}}tvrtl{\'i}k}, \citenamefont {Toma{\v{s}}t{\'i}k}, \citenamefont {Bekta{\c{s}}}, \citenamefont {Svoboda},\ and\ \citenamefont {Bange}}]{Simurka:2018}%
  \BibitemOpen
  \bibfield  {author} {\bibinfo {author} {\bibfnamefont {L.}~\bibnamefont {{\v{S}}imurka}}, \bibinfo {author} {\bibfnamefont {R.}~\bibnamefont {{\v{C}}tvrtl{\'i}k}}, \bibinfo {author} {\bibfnamefont {J.}~\bibnamefont {Toma{\v{s}}t{\'i}k}}, \bibinfo {author} {\bibfnamefont {G.}~\bibnamefont {Bekta{\c{s}}}}, \bibinfo {author} {\bibfnamefont {J.}~\bibnamefont {Svoboda}},\ and\ \bibinfo {author} {\bibfnamefont {K.}~\bibnamefont {Bange}},\ }\href {https://doi.org/10.1007/s11696-018-0420-z} {\bibfield  {journal} {\bibinfo  {journal} {Chemical Papers}\ }\textbf {\bibinfo {volume} {72}},\ \bibinfo {pages} {2143} (\bibinfo {year} {2018})}\BibitemShut {NoStop}%
\bibitem [{\citenamefont {Dingemans}\ \emph {et~al.}(2011)\citenamefont {Dingemans}, \citenamefont {Helvoirt}, \citenamefont {de~Sanden},\ and\ \citenamefont {Kessels}}]{Dingemans:2011}%
  \BibitemOpen
  \bibfield  {author} {\bibinfo {author} {\bibfnamefont {G.}~\bibnamefont {Dingemans}}, \bibinfo {author} {\bibfnamefont {C.~V.}\ \bibnamefont {Helvoirt}}, \bibinfo {author} {\bibfnamefont {M.~V.}\ \bibnamefont {de~Sanden}},\ and\ \bibinfo {author} {\bibfnamefont {W.~M.}\ \bibnamefont {Kessels}},\ }\href {https://doi.org/10.1149/1.3572283} {\bibfield  {journal} {\bibinfo  {journal} {ECS Transactions}\ }\textbf {\bibinfo {volume} {35}},\ \bibinfo {pages} {191} (\bibinfo {year} {2011})}\BibitemShut {NoStop}%
\bibitem [{\citenamefont {Taniguchi}\ \emph {et~al.}(1990)\citenamefont {Taniguchi}, \citenamefont {Tanaka}, \citenamefont {Hamaguchi},\ and\ \citenamefont {Imai}}]{Taniguchi:1990}%
  \BibitemOpen
  \bibfield  {author} {\bibinfo {author} {\bibfnamefont {K.}~\bibnamefont {Taniguchi}}, \bibinfo {author} {\bibfnamefont {M.}~\bibnamefont {Tanaka}}, \bibinfo {author} {\bibfnamefont {C.}~\bibnamefont {Hamaguchi}},\ and\ \bibinfo {author} {\bibfnamefont {K.}~\bibnamefont {Imai}},\ }\href {https://doi.org/10.1063/1.345563} {\bibfield  {journal} {\bibinfo  {journal} {Journal of Applied Physics}\ }\textbf {\bibinfo {volume} {67}},\ \bibinfo {pages} {2195} (\bibinfo {year} {1990})}\BibitemShut {NoStop}%
\bibitem [{\citenamefont {Gorodetsky}\ \emph {et~al.}(1996)\citenamefont {Gorodetsky}, \citenamefont {Savchenkov},\ and\ \citenamefont {Ilchenko}}]{Gorodetsky:1996}%
  \BibitemOpen
  \bibfield  {author} {\bibinfo {author} {\bibfnamefont {M.~L.}\ \bibnamefont {Gorodetsky}}, \bibinfo {author} {\bibfnamefont {A.~A.}\ \bibnamefont {Savchenkov}},\ and\ \bibinfo {author} {\bibfnamefont {V.~S.}\ \bibnamefont {Ilchenko}},\ }\href {https://doi.org/10.1364/OL.21.000453} {\bibfield  {journal} {\bibinfo  {journal} {Opt. Lett.}\ }\textbf {\bibinfo {volume} {21}},\ \bibinfo {pages} {453} (\bibinfo {year} {1996})}\BibitemShut {NoStop}%
\end{thebibliography}%

\end{document}